\documentclass[journal]{IEEEtran}
\ifCLASSINFOpdf
\else
\fi
\usepackage{booktabs} 
\usepackage{graphicx}
\usepackage{amsmath}
\usepackage{amssymb}
\usepackage{graphicx} 
\usepackage{subfigure}
\usepackage{graphicx}
\usepackage{algorithm}
\usepackage{algorithmic}
\usepackage{multirow}
\usepackage{makecell}
\usepackage{array}
\usepackage{url}

\newcolumntype{I}{!{\vrule width 3pt}}
\newlength\savedwidth
\newcommand\whline{\noalign{\global\savedwidth\arrayrulewidth
                            \global\arrayrulewidth 1.5pt}%
                   \hline
                   \noalign{\global\arrayrulewidth\savedwidth}}

\newlength\savewidth
\newcommand\shline{\noalign{\global\savewidth\arrayrulewidth
                            \global\arrayrulewidth 1pt}%
                   \hline
                   \noalign{\global\arrayrulewidth\savewidth}}

\newcommand{\tabincell}[2]{\begin{tabular}{@{}#1@{}}#2\end{tabular}}

\def\a{{\bf a}}
\def\B{{\bf B}}
\def\bb{{\bf b}}
\def\C{{\bf C}}

\def\h{{\bf h}}

\def\X{{\bf X}}

\def\p{{\bf p}}
\def\Q{{\bf Q}}

\def\S{{\bf S}}
\def\x{{\bf x}}

\def\z{{\bf z}}

\def\u{{\bf u}}

\def\X{{\bf X}}
\def\z{{\bf z}}

\def\0{{\bf 0}}
\def\1{{\bf 1}}

\def\RB{{\mathbb R}}
\def\IB{{\mathbb I}}

\def\LM{{\mathcal L}}

\begin{document}
%
\title{Deep Discrete Supervised Hashing}
%
%
%


\author{Qing-Yuan Jiang,~Xue Cui~and~Wu-Jun Li,~\IEEEmembership{Member,~IEEE}
\thanks{All authors are with the National Key Laboratory for Novel Software
Technology, Department of Computer Science and Technology, Nanjing University, Nanjing 210023, China. Wu-Jun Li is the corresponding author.
E-mail: \{jiangqy,~cuix\}@lamda.nju.edu.cn; liwujun@nju.edu.cn}}

%
%

\markboth{}
{Shell \MakeLowercase{\textit{et al.}}: Bare Demo of IEEEtran.cls for IEEE Journals}
%



\maketitle

\begin{abstract}
Hashing has been widely used for large-scale search due to its low storage cost and fast query speed. By using supervised information, supervised hashing can significantly outperform unsupervised hashing. Recently, discrete supervised hashing and deep hashing are two representative progresses in supervised hashing. On one hand, hashing is essentially a discrete optimization problem. Hence, utilizing supervised information to directly guide discrete~(binary) coding procedure can avoid sub-optimal solution and improve the accuracy. On the other hand, deep hashing, which integrates deep feature learning and hash-code learning into an end-to-end architecture, can enhance the feedback between feature learning and hash-code learning. The key in discrete supervised hashing is to adopt supervised information to directly guide the discrete coding procedure in hashing. The key in deep hashing is to adopt the supervised information to directly guide the deep feature learning procedure. However, there have not existed works which can use the supervised information to \emph{directly} guide both discrete coding procedure and deep feature learning procedure in the same framework. In this paper, we propose a novel deep hashing method, called deep discrete supervised hashing~(DDSH), to address this problem. DDSH is the first deep hashing method which can utilize supervised information to \emph{directly} guide both discrete coding procedure and deep feature learning procedure, and thus enhance the feedback between these two important procedures. Experiments on three real datasets show that DDSH can outperform other state-of-the-art baselines, including both discrete hashing and deep hashing baselines, for image retrieval.

\end{abstract}

\begin{IEEEkeywords}
Image retrieval, deep learning, deep supervised hashing.
\end{IEEEkeywords}

%
\IEEEpeerreviewmaketitle

\section{Introduction}
%
%
%
%
\IEEEPARstart{D}ue to the explosive increasing of data in real applications, nearest neighbor~(NN)~\cite{DBLP:conf/mfcs/Andoni11} search plays a fundamental role in many areas including image retrieval, computer vision, machine learning and data mining. In many real applications, there is no need to return the exact nearest neighbors for every given query and approximate nearest neighbor~(ANN) is enough to achieve satisfactory search~(retrieval) performance. Hence ANN search has attracted much attention in recent years~\cite{DBLP:conf/focs/AndoniI06,DBLP:conf/sigir/ZhangWCL10,DBLP:conf/cvpr/HeWS13,DBLP:journals/tip/ShenZ0SST16,DBLP:conf/sigir/ZhangSLHLC16}.

Over the last decades, hashing has become an active sub-area of ANN search~\cite{DBLP:journals/tip/ShenZ0SST16,DBLP:conf/nips/KongL12,DBLP:journals/tip/ZhangLZZZ15}. The goal of hashing is to map the data points to binary codes with hash functions which can preserve the similarity in the original space of the data points. With the binary hash code representation, the storage cost for the data points can be dramatically reduced. Furthermore, hashing based ANN search is able to achieve a constant or sub-linear time complexity~\cite{DBLP:conf/cvpr/LiuWJJC12}. Hence, hashing has become a promising choice for efficient ANN search in large-scale datasets because of its low storage cost and fast query speed~\cite{DBLP:conf/vldb/GionisIM99,DBLP:conf/focs/AndoniI06,DBLP:conf/iccv/KulisG09,DBLP:conf/sigir/ZhangWCL10,DBLP:conf/cvpr/WangKC10,DBLP:conf/cvpr/HeoLHCY12,DBLP:conf/cvpr/LiuWJJC12,DBLP:conf/cvpr/HeWS13,DBLP:conf/sigir/ZhangSLHLC16,DBLP:journals/tip/GuoDLHS17}.

Existing hashing methods can be divided into two main categories: data-independent methods and data-dependent methods. Data-independent hashing methods usually adopt random projections as hash functions to map the data points from the original space into a Hamming space of binary codes. That is to say, these methods do not use any training data to learn hash functions and binary codes. Representative data-independent hashing methods include locality-sensitive hashing~(LSH)~\cite{DBLP:conf/focs/AndoniI06,DBLP:conf/compgeom/DatarIIM04}, kernelized locality-sensitive hashing~(KLSH)~\cite{DBLP:conf/iccv/KulisG09}. Typically, data-independent hashing methods need long binary code to achieve satisfactory retrieval performance. Data-dependent hashing methods, which are also called \emph{learning to hash} methods, try to learn the hash functions from data. Recent works~\cite{DBLP:conf/cvpr/GongL11,DBLP:conf/cvpr/HeoLHCY12,DBLP:conf/cvpr/LiuWJJC12,DBLP:conf/cvpr/ShenSSHT13,DBLP:conf/iccv/LinSSH13,DBLP:conf/cvpr/LinSSHS14,DBLP:conf/iccv/SongLJMS15,DBLP:conf/cvpr/ShenSLS15} have shown that data-dependent methods can achieve comparable or even better accuracy with shorter binary hash codes, compared with data-independent methods. Therefore, data-dependent methods have received more and more attention.

Existing data-dependent hashing methods can be further divided into unsupervised hashing methods and supervised hashing methods, based on whether supervised information is used for learning or not. Unsupervised hashing methods aim to preserve the metric~(Euclidean neighbor) structure among the training data. Representative unsupervised hashing methods include spectral hashing~(SH)~\cite{DBLP:conf/nips/WeissTF08}, iterative quantization~(ITQ)~\cite{DBLP:conf/cvpr/GongL11}, isotropic hashing~(IsoHash)~\cite{DBLP:conf/nips/KongL12}, spherical hashing~(SPH)~\cite{DBLP:conf/cvpr/HeoLHCY12}, inductive manifold hashing~(IMH)~\cite{DBLP:conf/cvpr/ShenSSHT13}, anchor graph hashing~(AGH)~\cite{DBLP:conf/icml/LiuWKC11}, discrete graph hashing~(DGH)~\cite{DBLP:conf/nips/LiuMKC14}, latent semantic minimal hashing~(LSMH)~\cite{DBLP:journals/tip/LuZL17} and global hashing system~(GHS)~\cite{DBLP:journals/tip/TianT17}. Due to the semantic gap~\cite{smeulders2000content}, unsupervised hashing methods usually can not achieve satisfactory retrieval performance in real applications. Unlike unsupervised hashing methods, supervised hashing methods aim to embed the data points from the original space into the Hamming space by utilizing supervised information to facilitate hash function learning or hash-code learning. Representative supervised hashing methods include semantic hashing~\cite{DBLP:journals/ijar/SalakhutdinovH09}, self-taught hashing~(STH)~\cite{DBLP:conf/sigir/ZhangWCL10}, supervised hashing with kernels~(KSH)~\cite{DBLP:conf/cvpr/LiuWJJC12}, latent factor hashing~(LFH)~\cite{DBLP:conf/sigir/ZhangZLG14}, fast supervised hashing~(FastH)~\cite{DBLP:conf/cvpr/LinSSHS14}, supervised discrete hashing~(SDH)~\cite{DBLP:conf/cvpr/ShenSLS15} and column sampling based discrete supervised hashing~(COSDISH)~\cite{DBLP:conf/aaai/KangLZ16}. By using supervised information for learning, supervised hashing can significantly outperform unsupervised hashing in real applications such as image retrieval.


Hashing is essentially a discrete optimization problem. Because it is difficult to directly solve the discrete optimization problem, early hashing methods~\cite{DBLP:conf/cvpr/WangKC10,DBLP:conf/cvpr/GongL11,DBLP:conf/iccv/LinSSH13} adopt relaxation strategies to solve a proximate continuous problem which might result in a sub-optimal solution. Specifically, relaxation based hashing methods utilize supervised information to guide continuous hash code learning at the first stage. Then they convert continuous hash code into binary code by using rounding technology at the second stage. Recently, several discrete hashing methods, e.g., DGH~\cite{DBLP:conf/nips/LiuMKC14}, FastH~\cite{DBLP:conf/cvpr/LinSSHS14}, SDH~\cite{DBLP:conf/cvpr/ShenSLS15} and COSDISH~\cite{DBLP:conf/aaai/KangLZ16}, which try to directly learn the discrete binary hash codes, have been proposed with promising performance. In particular, several discrete supervised hashing methods, including FastH~\cite{DBLP:conf/cvpr/LinSSHS14}, SDH~\cite{DBLP:conf/cvpr/ShenSLS15} and COSDISH~\cite{DBLP:conf/aaai/KangLZ16}, have shown better performance than traditional relaxation-based continuous hashing methods. The key in discrete supervised hashing is to adopt supervised information to \emph{directly} guide the discrete coding procedure, i.e., the discrete binary code learning procedure.

Most existing supervised hashing methods, including those introduced above, are based on hand-crafted features. One major shortcoming for these methods is that they cannot perform feature learning. That is, these hand-crafted features might not be optimally compatible with the hash code learning procedure. Hence, besides the progress about discrete hashing, there has appeared another progress in supervised hashing, which is called deep hashing~\cite{DBLP:conf/aaai/XiaPLLY14,DBLP:conf/cvpr/LaiPLY15,DBLP:journals/tip/ZhangLZZZ15,DBLP:conf/cvpr/ZhaoHWT15,zhuang2016fast,liudeep,DBLP:conf/aaai/CaoL0ZW16,DBLP:conf/ijcai/LiWK16,DBLP:conf/aaai/ZhuL0C16}. Representative deep hashing includes convolutional neural network hashing~(CNNH)~\cite{DBLP:conf/aaai/XiaPLLY14}, network in network hashing~(NINH)~\cite{DBLP:conf/cvpr/LaiPLY15}, deep semantic ranking hashing~(DSRH)~\cite{DBLP:conf/cvpr/ZhaoHWT15}, deep similarity comparison hashing~(DSCH)~\cite{DBLP:journals/tip/ZhangLZZZ15}, deep pairwise-supervised hashing~(DPSH)~\cite{DBLP:conf/ijcai/LiWK16}, deep hashing network~(DHN)~\cite{DBLP:conf/aaai/CaoL0ZW16}, deep supervised hashing~(DSH)~\cite{liudeep}, and deep quantization network~(DQN)~\cite{DBLP:conf/aaai/CaoL0ZW16}. Deep hashing adopts deep learning~\cite{DBLP:conf/nips/KrizhevskySH12,DBLP:conf/nips/CunBDHHHJ89} for supervised hashing. In particular, most deep hashing methods adopt deep feature learning to learn a feature representation for hashing. Many deep hashing methods integrate deep feature representation learning and hashing code learning into an end-to-end architecture. Under this architecture, feature learning procedure and hash-code learning procedure can give feedback to each other in learning procedure. Many works~\cite{DBLP:conf/ijcai/LiWK16,liudeep} have shown that using the supervised information to \emph{directly} guide the deep feature learning procedure can achieve better performance than other strategies~\cite{DBLP:conf/aaai/XiaPLLY14} which do not use supervised information to directly guide the deep feature learning procedure. Hence, the key in deep hashing is to adopt the supervised information to directly guide the deep feature learning procedure.


Both discrete supervised hashing and deep hashing have achieved promising performance in many real applications. However, there have not existed works which can use the supervised information to \emph{directly} guide both discrete~(binary) coding procedure and deep feature learning procedure in the same framework. In this paper, we propose a novel deep hashing method, called \underline{d}eep \underline{d}iscrete \underline{s}upervised \underline{h}ashing~(DDSH), to address this problem. The main contributions of DDSH are outlined as follows:
\begin{itemize}
     \item DDSH is the first deep hashing method which can utilize supervised information to \emph{directly} guide both discrete coding procedure and deep feature learning procedure.
     \item By integrating the discrete coding procedure and deep feature learning procedure into the same end-to-end framework, these two important procedures can give feedback to each other to make the hash codes and feature representation more compatible.
     \item  Experiments on three real datasets show that our proposed DDSH can outperform other state-of-the-art baselines, including both discrete hashing and deep hashing baselines.
\end{itemize}

The rest of this paper is organized as follows.
In Section~\ref{sec:NoPD},
we briefly introduce the notations and problem definition in this paper.
Then we describe DDSH in Section~\ref{sec:DDSH}.
We discuss the difference between DDSH and existing deep hashing methods in Section~\ref{sec:related}.
In Section~\ref{sec:exp}, we evaluate DDSH on three datasets
by carrying out the Hamming ranking task and hash lookup task.
Finally, we conclude the paper in Section~\ref{sec:conc}.

\section{Notation and Problem Definition}
\label{sec:NoPD}
\subsection{Notation}
 Some representative notations we use in this paper are outlined in Table~\ref{tab:notations}. More specifically, we use boldface uppercase letters like $\B$ to denote matrices. We use boldface lowercase letters like $\bb$ to denote vectors. The $(i,j)$th element in matrix $\B$ is denoted as $B_{ij}$. $\B^T$ is the transpose of $\B$ and $\Vert \bb\Vert_2$ is the Euclidean norm of vector $\bb$. We use the capital Greek letter like $\Omega$ to denote the set of indices. We use the symbol $\bullet$ to denote the Hadamard product~(i.e., element-wise product). The square of a vector~(or a matrix) like $\bb^2$ indicates element-wise square, i.e., $\bb^2=\bb\bullet\bb$.

\begin{table}[t]
\centering
\caption{Notation.}
\label{tab:notations}
\begin{tabular}{|c||c|}
 \hline
Notation & Meaning \\
\hline
$\B$ & boldface uppercase, matrix\\\hline
$\bb$ & boldface lowercase, vector \\\hline
$B_{ij}$ & the $(i,j)$th element in matrix $\B$\\\hline
$\B^T$ & transpose of matrix $\B$\\\hline
$\Vert\bb\Vert_2$ &  Euclidean norm of vector $\bb$\\\hline
$\Omega$ &  capital Greek letter, set of indices\\\hline
$\bullet$ &  Hadamard product~(i.e., element-wise product) \\\hline
$\bb^2$ & element-wise square of vector, i.e., $\bb^2=\bb\bullet\bb$\\
\hline
 \end{tabular}
\end{table}

\subsection{Problem Definition}
\label{sec:pd}
Although supervised information can also be triplet labels~\cite{DBLP:conf/cvpr/LaiPLY15,DBLP:journals/tip/ZhangLZZZ15,DBLP:conf/cvpr/ZhaoHWT15,zhuang2016fast} or pointwise labels~\cite{DBLP:conf/cvpr/ShenSLS15}, in this paper we only focus on the setting with pairwise labels~\cite{DBLP:conf/aaai/XiaPLLY14,DBLP:conf/aaai/ZhuL0C16,DBLP:conf/ijcai/LiWK16,DBLP:conf/aaai/CaoL0ZW16,liudeep} which is a popular setting in supervised hashing. The technique in this paper can also be adapted to settings with triplet labels, which will be pursued in our future work.

We use $\X=\{\x_i\}_{i=1}^n$ to denote a set of training points. In supervised hashing with pairwise labels, the supervised information $\S=\{-1,1\}^{n\times n}$ between data points is also available for training procedure, where $S_{ij}$ is defined as follows:
\begin{align}
S_{ij}=\left\{
\begin{aligned}
 1, &\; \text{$\x_i$ and $\x_j$ are similar.}\\
-1, &\;\text{otherwise}.
\end{aligned}
\right.\nonumber
\end{align}

Supervised hashing aims at learning a hash function to  map the data points from the original space into the binary code space~(or called Hamming space), with the semantic~(supervised) similarity in $\S$ preserved in the binary code space. We define the hash function as: $h(\x)\in\{-1,+1\}^c,\forall\x\in\X$, where $c$ is the binary code length. The Hamming distance between binary codes $\bb_i=h(\x_i)$ and $\bb_j=h(\x_j)$ is defined as follows:
\begin{align}
\text{dist}_H(\bb_i,\bb_j)=\frac{1}{2}(c-\bb_i^T\bb_j).\nonumber
\end{align}
To preserve the similarity between data points, the Hamming distance between the binary codes $\bb_i=h(\x_i)$ and $\bb_j=h(\x_j)$ should be relatively small if the data points $\x_i$ and $\x_j$ are similar, i.e., $S_{ij}=1$. On the contrary, the Hamming distance between the binary codes $\bb_i=h(\x_i)$ and $\bb_j=h(\x_j)$ should be relatively large if the data points $\x_i$ and $\x_j$ are dissimilar, i.e., $S_{ij}=-1$.
In other words, the goal of supervised hashing is to solve the following problem:
\begin{align}
\min_{h}\;\LM(h)=&\sum_{i,j=1}^n L(h(\x_i),h(\x_j);S_{ij})\nonumber\\
=&\sum_{i,j=1}^n L(\bb_i,\bb_j;S_{ij}),
\label{pb:ori}
\end{align}
where $L(\cdot)$ is a loss function.

There have appeared various loss functions in supervised hashing. For example, KSH~\cite{DBLP:conf/cvpr/LiuWJJC12} uses $L_2$ function, which is defined as follows:
\begin{align}
L(\bb_i,\bb_j;S_{ij})=(S_{ij}-\frac{1}{c}\sum_{m=1}^c b_i^m b_j^m)^2. \nonumber
\end{align}
where $b_i^m$ is the $m$th element in vector $\bb_i$. Please note that our DDSH is able to adopt many different kinds of loss functions. In this paper, we only use $L_{2}$ loss function as an example, and leave other loss functions for further study in future work.



\begin{figure*}[t]
\begin{center}
\includegraphics[scale=0.26]{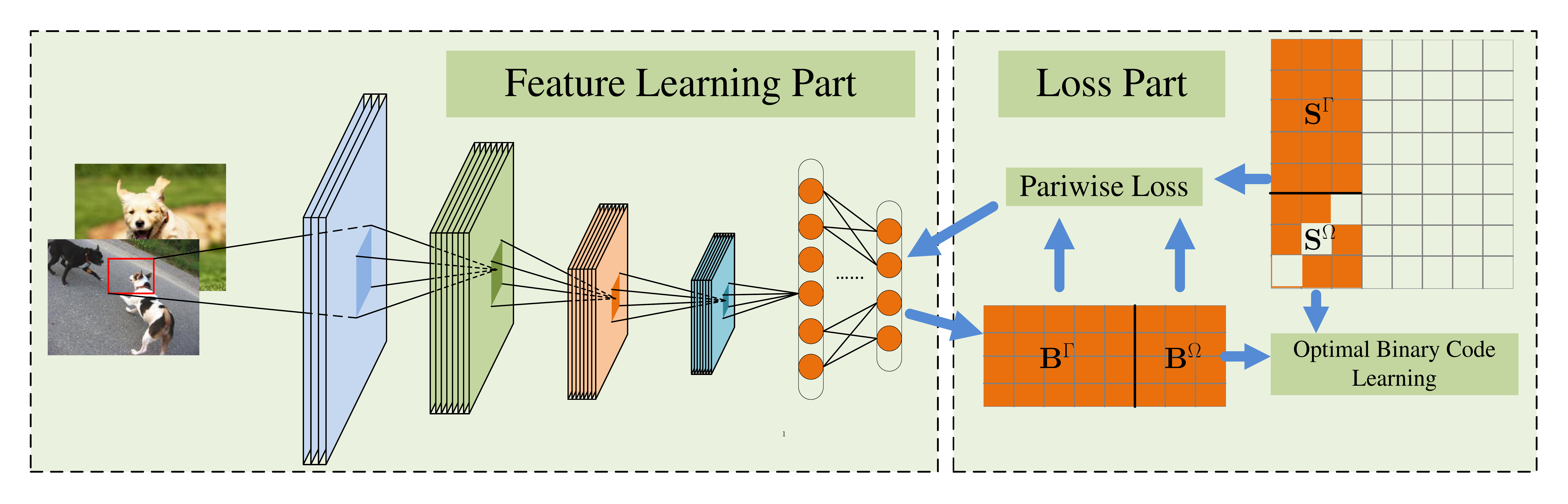}
\end{center}
\caption{The model architecture of DDSH. DDSH is an end-to-end deep learning framework which consists of two main components: loss part and feature learning part. The loss part contains the discrete coding procedure~(to learn the binary codes $\B^\Omega$), and the feature learning part contains the deep feature learning procedure~(to learn the $F(\x;\Theta)$ for $\x$ indexed by $\Gamma$). During each iteration, we adopt an alternating strategy to learn binary codes and feature representation alternatively, both of which are directly guided by supervised information. Hence, DDSH can enhance the feedback between the discrete coding procedure and the deep feature learning procedure.}
\label{fig:framework}
\end{figure*}

\section{Deep Discrete Supervised Hashing}
\label{sec:DDSH}

In this section, we describe the details of DDSH, including the model architecture and learning algorithm.

\subsection{Model Architecture}
DDSH is an end-to-end deep hashing method which is able to simultaneously perform feature learning and hash code learning in the same framework. The model architecture of DDSH is shown in Figure~\ref{fig:framework}, which contains two important components: \emph{loss part} and \emph{feature learning part}. The loss part contains the discrete coding procedure which aims to learn optimal binary code to preserve semantic pairwise similarity. The feature learning part contains the deep feature learning procedure which tries to learn a compatible deep neural network to extract deep representation from scratch. For DDSH, discrete coding procedure and deep feature learning are integrated into an end-to-end framework. More importantly, both procedures are \emph{directly} guided by supervised information.

\subsubsection{Loss Part}
Inspired by COSDISH~\cite{DBLP:conf/aaai/KangLZ16}, we use column-sampling method to
split the whole training set into two parts.
More specifically, we randomly sample a subset $\Omega$ of $\Phi=\{1,2,\dots,n\}$
and generate $\Gamma=\Phi\setminus\Omega$~($\vert\Gamma\vert\gg\vert\Omega\vert$).
Then we split the whole training set $\X$ into two subsets $\X^{\Omega}$ and $\X^{\Gamma}$,
where $\X^{\Omega}$ and $\X^{\Gamma}$ denote the training data points indexed by $\Omega$ and $\Gamma$ respectively.

Similarly, we sample $\vert\Omega\vert$ columns of $\S$ with the corresponding sampled columns indexed by $\Omega$.
Then, we approximate the original problem in~(\ref{pb:ori}) by only using the sampled columns of $\S$:
\begin{align}
\min_{h}\;\LM(h)=&\sum_{i\in\Omega}\sum_{j=1}^{n}L(h(\x_i),h(\x_j);S_{ij})\nonumber\\
=&\sum_{\x_i\in{\X^\Omega}}\sum_{\x_j\in{\X^\Gamma}} L(h(\x_i),h(\x_j);S_{ij})\nonumber\\
&+\sum_{\x_i,\x_j\in{\X^\Omega}} L(h(\x_i),h(\x_j);S_{ij}).
\label{optobj}
\end{align}

Then we introduce auxiliary variables to solve problem~(\ref{optobj}).
More specifically, we utilize auxiliary variables $\B^\Omega =\{\bb_i|i\in\Omega\}$ with $\bb_i \in \{-1,+1\}^{c}$ to replace part of the binary codes generated by the hash function, i.e., $h(\X^{\Omega})$. Here, $h(\X^{\Omega}) = \{h(\x_i)|\x_i\in\X^\Omega\}$.
Then we rewrite the problem~(\ref{optobj}) as follows:
\begin{align}
\min_{h,\B^{\Omega}}\;\LM(h,\B^{\Omega})=&\sum_{i\in\Omega}\sum_{\x_j\in{\X^\Gamma}} L(\bb_i,h(\x_j);S_{ij})\nonumber\\
&+\sum_{i,j\in{\Omega}} L(\bb_i,\bb_j;S_{ij})\nonumber\\
\text{s.t.}\;&\bb_i\in\{-1,+1\}^{c}, \forall i\in \Omega
\label{optobj1}
\end{align}

The problem in~(\ref{optobj1}) is the final loss function~(objective) to learn by DDSH. We can find that the whole training set is divided into two subsets $\X^\Omega$ and $\X^\Gamma$. The binary codes of $\X^\Omega$, i.e., $\B^\Omega$, are directly learned from the objective function in~(\ref{optobj1}), but the binary codes of $\X^\Gamma$ are generated by the hash function $h(\cdot)$. $h(\cdot)$ is defined based on the output of the deep feature learning part, which will be introduced in the following subsection.

The learning of $\B^\Omega$ contains the discrete coding procedure, which is directly guided by the supervised information. The learning of $h(\cdot)$ contains the deep feature learning procedure, which is also directly guided by the supervised information. Hence, our DDSH can utilize supervised information to directly guide both discrete coding procedure and deep feature learning procedure in the same end-to-end deep framework. This is different from existing deep hashing methods which either use relaxation strategy without discrete coding or do not use the supervised information to directly guide the discrete coding procedure.

Please note that ``\emph{directly guided}'' in this paper means that the supervised information is directly included in the corresponding terms in the loss function. For example, the supervised information $S_{ij}$ is directly included in all terms about the discrete codes $\B^{\Omega}$ in~(\ref{optobj1}), which means that the discrete coding procedure is directly guided by the supervised information. Furthermore, the supervised information $S_{ij}$ is also directly included in the term about the deep feature learning function $h(\x_j)$ in~(\ref{optobj1}), which means that the deep feature learning procedure is also directly guided by the supervised information. To the best of our knowledge, DDSH is the first deep hashing method which can utilize supervised information to directly guide both discrete coding procedure and deep feature learning procedure, and thus enhance the feedback between these two important procedures.

\subsubsection{Feature Learning Part}

The binary codes of $\X^\Gamma$ are generated by the hash function $h(\cdot)$, which is defined based on the output of the deep feature learning part. More specifically, we define our hash function as: $h(\x)=\text{sign}(F(\x;\Theta))$, where $\text{sign}(\cdot)$ is the element-wise sign function. $F(\x;\Theta)$ denotes the output of the feature learning part and $\Theta$ denotes all parameters of the deep neural network.

We adopt a convolutional neural network~(CNN) from~\cite{DBLP:conf/bmvc/ChatfieldSVZ14}, {i.e., CNN-F,} as our deep feature learning part. We replace the last layer of CNN-F as one fully-connected layer to project the output of the second last layer to $\RB^c$ space. More specifically, the feature learning part contains 5 convolutional layers~(``conv1-conv5'') and 3 fully-connected layers~(``full6-full8''). The detailed configuration of the 5 convolutional layers is shown in Table~\ref{tab:conf_conv}. In Table~\ref{tab:conf_conv}, ``filter size'' denotes the number of convolutional filters and their receptive field size. ``stride'' specifies the convolutional stride. ``pad'' indicates the number of pixels to add to each size of the input. ``LRN'' denotes whether Local Response Normalization~(LRN)~\cite{DBLP:conf/nips/KrizhevskySH12} is applied or not. ``pool'' denotes the down-sampling factor. The detailed configuration of the 3 fully-connected layers is shown in Table~\ref{tab:conv_full}, where the ``configuration" shows the number of nodes in each layer.
\begin{table}[htb]
\centering
\caption{Configuration of the convolutional layers in DDSH.}
\label{tab:conf_conv}
\begin{tabular}{|c||c|c|c|c|c|}
 \hline
 \multirow{2}{*}{Layer} &
 \multicolumn{5}{c|}{{Configuration}}\\
 \cline{2-6} & filter size & stride & pad & LRN & pool\\
 \hline \hline
conv1 & $64\times 11\times 11$ & $4\times 4$  & {0} & yes & $2\times 2$\\
\hline
conv2 & $256\times 5\times 5$ & $1\times 1$  & {2} & yes & $2\times 2$\\
\hline
conv3 & $256\times 3\times 3$ & $1\times 1$  & {1} & no & -\\
\hline
conv4 & $256\times 3\times 3$ & $1\times 1$  & {1} & no & -\\
\hline
conv5 & $256\times 3\times 3$ & $1\times 1$  & {1} & no & $2\times 2$\\
\hline
 \end{tabular}
\end{table}

\begin{table}[htb]
\centering
\caption{Configuration of the fully-connected layers in DDSH.}
\label{tab:conv_full}
\begin{tabular}{|c||c|}
 \hline
Layer & Configuration \\
\hline\hline
full6 & 4096\\\hline
full7 & 4096\\\hline
full8 & hash code length $c$\\\hline
 \end{tabular}
\end{table}

We adopt the Rectified Linear Unit (ReLU)~\cite{DBLP:conf/nips/KrizhevskySH12} as activation function for all the first seven layers. For the last layer, we utilize identity function as the activation function.
\subsection{Learning}
After randomly sampling $\Omega$ at each iteration, we utilize an alternating learning strategy to solve problem~(\ref{optobj1}).

More specifically, each time we learn one of the variables $\B^{\Omega}$ and $h(F(\x;\Theta))$ with the other fixed. When $h(F(\x;\Theta))$ is fixed, we directly learn the discrete hash code $\B^{\Omega}$ over binary variables. When $\B^{\Omega}$ is fixed, we update the parameter $\Theta$ of the deep neural network.

\subsubsection{Learn $\B^{\Omega}$ with $h(F(\x;\Theta))$ Fixed}

When $h(F(\x;\Theta))$ is fixed, it's easy to transform problem~(\ref{optobj1}) into a binary quadratic programming~(BQP) problem as that in TSH~\cite{DBLP:conf/iccv/LinSSH13}. Each time we optimize one bit for all points. Then, the optimization of the $k$th bit of all points in $\B^\Omega$ is given by:
\begin{align}
\min_{\bb^k}\;&[\bb^k]^T\Q^k\bb^k+[\bb^k]^T\p^k\nonumber\\
\text{s.t.}\;&\bb^k\in\{-1,+1\}^{\vert\Omega\vert}
\label{optobjbk}
\end{align}
where $\bb^k$ denotes the $k$th column of $\B^\Omega$, and
\begin{align}
Q^k_{\substack{ij\\i\neq j}}=&-2(cS_{ij}^{\Omega}-\sum_{m=1}^{k-1}b_i^mb_j^m)\nonumber\\
Q^k_{ii}=&0\nonumber\\
p^k_{i}=&-2\sum_{l=1}^{\vert\Gamma\vert}
B_{lk}^\Gamma(cS^{\Gamma}_{li}-\sum_{m=1}^{k-1}B_{lm}^\Gamma b_{i}^{m}). \nonumber
\end{align}
Here, $b_i^m$ denotes the $m$th bit of $\bb_i$ and $p^k_i$ denotes the $i$th element of $\p^k$.

Following COSDISH, we can easily solve problem~(\ref{optobjbk}) by transforming the BQP problem into an equally clustering problem~\cite{yang2013new}.

\subsubsection{Learn $h(F(\x;\Theta))$ with $\B^{\Omega}$ Fixed}
Because the derivative of the hash function $h(\x)=\text{sign}(F(\x;\Theta))$ is 0 everywhere except at 0,
we cannot use back-propagation~(BP) methods to update the neural network parameters. So we relax $\text{sign}(\cdot)$ as $h(\x)=\text{tanh}(F(\x;\Theta))$  inspired by~\cite{DBLP:conf/iccv/SongLJMS15}. Then we optimize the following problem:
\begin{align}
\min_{h}\;\LM(h)=&\sum_{i\in{\Omega}}\sum_{\x_j\in{\X^\Gamma}} L(\bb_i,h(\x_j);S_{ij})\nonumber\\
\text{s.t.}\;&h(\x_j)=\text{tanh}(F(\x_j;\Theta))
\label{optobjh}
\end{align}

To learn the CNN parameter $\Theta$, we utilize a back-propagation algorithm.
That is, each time we sample a mini-batch of data points, and then use BP algorithm based on the sampled data.

We define the output of CNN as $\z_j=F(\x_j;\Theta)$  and $\a_j=\text{tanh}(\z_j)$. Then we can compute the gradient of $\a_j$ and $\z_j$ as follows:
\begin{align}
\frac{\partial \LM}{\partial \a_j}=&\sum_{i\in{\Omega}}\frac{\partial L(\bb_i,\a_j;S_{ij})}{\partial \a_j}\nonumber\\
=&\sum_{i\in {\Omega}}2(\a_j^T\bb_i-S_{ij})\bb_i
\label{optgradaj}
\end{align}
and
\begin{align}
\frac{\partial \LM}{\partial \z_j}=&\frac{\partial \LM}{\partial \a_j}\bullet(1-\a_j^2)\nonumber\\
=&\sum_{i\in{\Omega}}2(\a_j^T\bb_i-S_{ij})\bb_i\bullet(1-\a_j^2)
\label{optgrad}
\end{align}

Then, we can use chain rule to compute $\frac{\partial \LM}{\partial \Theta}$ based on $\frac{\partial \LM}{\partial \a_j}$ and $\frac{\partial \LM}{\partial \z_j}$.

We summarize the whole learning algorithm for DDSH in Algorithm~\ref{alg:Framwork}.

\begin{algorithm}[tb]
\caption{The learning algorithm for DDSH}
\label{alg:Framwork}
\begin{algorithmic}
\REQUIRE ~~\\
    Training set $\X$;\\
    Code length $c$; \\
    Supervised information $\S$.\\
\ENSURE ~~\\
    Parameter $\Theta$ of the deep neural network.

\STATE \textbf{Initialization}
\STATE Initialize neural network parameter $\Theta$, mini-batch size $M$ and iteration number $T_{out},T_{in}$
\STATE Initialize $\B = \{\bb_i|i=1,2,\cdots,n\}$
\FOR {$iter=1,2,\dots,T_{out}$}
    \STATE Randomly sample $\Omega$ and set $\Gamma=\Phi\setminus\Omega$
    \STATE Split training set $\X$ into $\X^{\Omega}$ and $\X^{\Gamma}$.
    \STATE Split $\B$ into $\B^{\Omega}$ and $\B^{\Gamma}$.
        \FOR {$epoch=1,2,\dots,T_{in}$}
            \FOR {$k=1,2,\dots,c$}
                \STATE Construct the BQP problem for the $k$th bit using~(\ref{optobjbk}).
                \STATE Construct the clustering problem to solve the BQP problem for the $k$th bit.
            \ENDFOR

            \FOR {$t=1,2,\dots,\vert\Gamma\vert/M$}
                \STATE Randomly sample $M$ data points from $\X^\Gamma$ to construct a mini-batch.
                \STATE Calculate $h(\x_j)$ for each data point $\x_j$ in the mini-batch by forward propagation.
                \STATE Calculate the gradient according to~(\ref{optgrad}).
                \STATE Update the parameter $\Theta$ by using back propagation.
                \STATE Update $\bb_j=\text{sign}(h(\x_j))$ for each data point $\x_j$ in the mini-batch.
            \ENDFOR
        \ENDFOR
\ENDFOR
\end{algorithmic}
\end{algorithm}

\subsection{Out-of-Sample Extension for Unseen Data Points}
After training our DDSH model, we can adopt the learned deep hashing framework to predict the binary code for any unseen data point during training.

More specifically, given any point $\x_q\notin\X$, we use the following formula to predict its binary code:
\begin{align}
\bb_q=h(\x_q)=\text{sign}(F(\x_q;\Theta)), \nonumber
\end{align}
where $\Theta$ is the deep neural network parameter learned by DDSH model.
\section{Comparison to Related Work}
\label{sec:related}
Although a lot of deep hashing methods have been proposed, none of these methods can utilize supervised information to directly guide both discrete coding procedure and deep feature learning procedure.

Existing deep hashing methods either use relaxation strategy without discrete coding or do not use the supervised information to directly guide the discrete coding procedure. For example, CNNH~\cite{DBLP:conf/aaai/XiaPLLY14} is a two-step method which adopts relaxation strategy to learn continuous code in the first stage and performs feature learning in the second stage. The feature learning procedure in CNNH is not directly guided by supervised information. NINH~\cite{DBLP:conf/cvpr/LaiPLY15}, DHN~\cite{DBLP:conf/aaai/CaoL0ZW16} and DSH~\cite{liudeep} adopt relaxation strategy to learn continuous code. DPSH~\cite{DBLP:conf/ijcai/LiWK16} and DQN~\cite{DBLP:conf/aaai/CaoL0ZW16} can learn binary code in the training procedure. However, DSPH and DQN do not utilize the supervised information to directly guide the discrete coding procedure. The objective function of DPSH can be written as: $\LM_{\text{DPSH}}=-\sum_{S_{ij}\in\S}(S_{ij}\Theta_{ij}-\log(1+e^{\Theta_{ij}}))+\eta\sum_{i=1}^{n}\Vert\bb_i-\u_i\Vert^2_F$~\footnote{For DPSH, supervised information $S_{ij}$ is defined on $\{0,1\}$.}, where $\Theta_{ij}=\frac{1}{2}\u_i^T\u_j$ and $\u_i$ denotes the output of the deep neural network. We can find that in DPSH the discrete coding procedure is not directly guided by supervised information, i.e., the supervised information is not directly included in the terms of $\{\bb_i\}$ in the objective function. The objective function of DQN can be written as: $\LM_{\text{DQN}}=\sum_{S_{ij}\in\S}(S_{ij}-\frac{\z_i^T\z_j}{\Vert\z_i\Vert\Vert\z_j\Vert})^2+\lambda\sum_{i=1}^n\Vert\z_i-\C\h_i\Vert^2_F$, where $\z_i$ denotes the output of the deep neural network and $\sum_{i=1}^n\Vert\z_i-\C\h_i\Vert^2_F$ denotes the product quantization loss. The discrete coding procedure is only contained in the term $\sum_{i=1}^n\Vert\z_i-\C\h_i\Vert^2_F$. We can find that in DQN the discrete coding procedure is not directly guided by supervised information either.

To the best of our knowledge, our DDSH is the first deep hashing method which can utilize supervised information to directly guide both discrete coding procedure and deep feature learning procedure in the same framework.

\section{Experiment}
\label{sec:exp}
We evaluate DDSH and other baselines on datasets from image retrieval applications. The open source deep learning library MatConvNet~\cite{DBLP:conf/mm/VedaldiL15} is used to implement our model. All experiments are performed on an NVIDIA K40 GPU server.

\subsection{Experimental Setting}
\subsubsection{Datasets}

We adopt three widely used image datasets to evaluate our proposed method. They are CIFAR-10\footnote {https://www.cs.toronto.edu/\textasciitilde kriz/cifar.html}~\cite{DBLP:conf/nips/KrizhevskySH12},
SVHN\footnote{http://ufldl.stanford.edu/housenumbers/}~\cite{netzer2011reading} and NUS-WIDE\footnote{http://lms.comp.nus.edu.sg/research/NUS-WIDE.htm}~\cite{DBLP:conf/civr/ChuaTHLLZ09}.

The CIFAR-10 dataset contains 60,000 images which are manually labeled into 10 classes including ``airplane'', ``automobile'', ``bird'', ``cat'', ``deer'', ``dog'', ``frog'', ``horse'', ``ship'' and ``truck''. It's a single-label dataset. The size of each image is 32$\times$32 pixels. Two images are treated as similar if they share the same label, i.e., they belong to the same class. Otherwise, they are considered to be dissimilar.

The SVHN dataset consists of 73,257 digits for training, 26,032 digits for testing and 531,131 additional samples. It is a real-world image dataset for recognizing digital numbers in natural scene images. The images are categorized into 10 classes, each corresponding to a digital number. SVHN is also a single-label dataset. Two images are treated as similar if they share the same label. Otherwise, they are considered to be dissimilar.

The NUS-WIDE dataset is a relatively large-scale image dataset which includes 269,648 images and the associated tags from Flickr website. It's a multi-label dataset where each image might be annotated with multi-labels. We select 186,577 data points that belong to the 10 most frequent concepts from the original dataset. Two images are treated as similar if they share at least one label. Otherwise, they are considered to be dissimilar.

Table~\ref{tab:example} illustrates some example points from the above three datasets.

\begin{table}[t]
\caption{Example points of the datasets.}\label{tab:example}
\centering
\begin{tabular}{|c|c|c|}
\hline
Dataset & Example & Label \\\hline\hline
\multirow{3}{*}{\vspace{-20pt}CIFAR-10} &
\begin{minipage}{.085\textwidth}\vspace{2pt}
\includegraphics[scale=0.85]{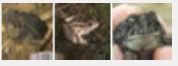}\vspace{2pt}
\end{minipage}
& ``frog''.
\\\cline{2-3}
&\begin{minipage}{.085\textwidth}\vspace{2pt}
\includegraphics[scale=0.85]{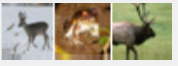}\vspace{2pt}
\end{minipage}
& ``deer''.
\\\cline{2-3}
& \begin{minipage}{.085\textwidth}\vspace{2pt}
\includegraphics[scale=0.85]{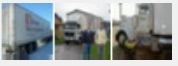}\vspace{2pt}
\end{minipage}
& ``truck''.
\\\hline
\multirow{3}{*}{\vspace{-20pt}SVHN} &
\begin{minipage}{.085\textwidth}\vspace{2pt}
\includegraphics[scale=0.85]{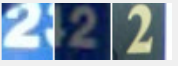}\vspace{2pt}
\end{minipage}
& ``2''.
\\\cline{2-3}
&\begin{minipage}{.085\textwidth}\vspace{2pt}
\includegraphics[scale=0.85]{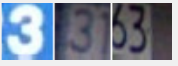}\vspace{2pt}
\end{minipage}
& ``3''.
\\\cline{2-3}
& \begin{minipage}{.085\textwidth}\vspace{2pt}
\includegraphics[scale=0.85]{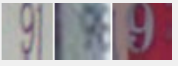}\vspace{2pt}
\end{minipage}
& ``9''.
\\\hline
\multirow{3}{*}{\vspace{-50pt}NUS-WIDE} &
\begin{minipage}{.15\textwidth}\vspace{2pt}
\includegraphics[scale=0.3]{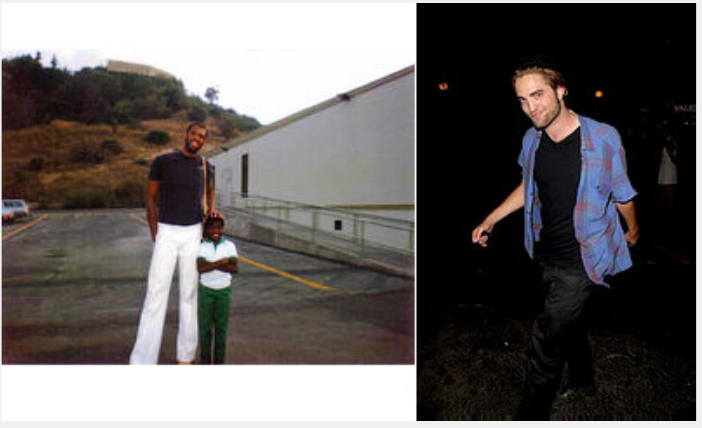}\vspace{2pt}
\end{minipage}
& ``person'', ``sky''.
\\\cline{2-3}
&\begin{minipage}{.085\textwidth}\vspace{4pt}
\includegraphics[scale=0.15]{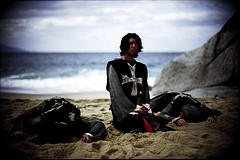}\vspace{4pt}
\end{minipage}
& \tabincell{c}{``clouds'', ``ocean'', \\ ``person'', ``sky'', ``water''. }
\\\cline{2-3}
&\begin{minipage}{.085\textwidth}\vspace{4pt}
\includegraphics[scale=0.15]{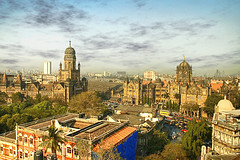}\vspace{4pt}
\end{minipage}
& \tabincell{c}{``road'', ``clouds'', \\ ``sky'', ``buildings''. }
\\\hline
\end{tabular}
\end{table}

For CIFAR-10 dataset, we randomly take 1,000 images~(100 images per class)
as query set and the remaining images as retrieval set.
For SVHN dataset, we randomly select 1,000 images~(100 images per class)
from testing set as query set and utilize the whole training set as retrieval set. For NUS-WIDE dataset, we randomly select 1,867 data points as query set
and the remaining data points as retrieval set.
For all datasets, we randomly select 5,000 data points from retrieval set as training set.

\subsubsection{Baselines and Evaluation Protocol}
We compare DDSH with nine state-of-the-art baselines, including LSH~\cite{DBLP:conf/compgeom/DatarIIM04}, ITQ~\cite{DBLP:conf/cvpr/GongL11}, LFH~\cite{DBLP:conf/sigir/ZhangZLG14}, FastH~\cite{DBLP:conf/cvpr/LinSSHS14}, SDH~\cite{DBLP:conf/cvpr/ShenSLS15}, COSDISH~\cite{DBLP:conf/aaai/KangLZ16}, DHN~\cite{DBLP:conf/aaai/ZhuL0C16}, DSH~\cite{liudeep}, and DPSH~\cite{DBLP:conf/ijcai/LiWK16}. These baselines are briefly introduced as follows:
\begin{itemize}
    \item Locality-sensitive hashing~(LSH)~\cite{DBLP:conf/compgeom/DatarIIM04}: LSH is a representative data-independent hashing method. LSH utilizes random projection to generate hash function.
    \item Iterative quantization~(ITQ)~\cite{DBLP:conf/cvpr/GongL11}: ITQ is a representative unsupervised hashing method. ITQ first projects data points into low space by utilizing principal component analysis~(PCA). Then ITQ minimizes the quantization error to learn binary code.
    \item Latent factor hashing~(LFH)~\cite{DBLP:conf/sigir/ZhangZLG14}: LFH is a supervised hashing method which tries to learn binary code based on latent factor models.
    \item Fast supervised hashing~(FastH)~\cite{DBLP:conf/cvpr/LinSSHS14}: FastH is supervised hashing method. FastH directly adopts graph-cut method to learn discrete binary code.
    \item Supervised discrete hashing~(SDH)~\cite{DBLP:conf/cvpr/ShenSLS15}: SDH is a pointwise supervised hashing method which utilizes the discrete cyclic coordinate descent~(DCC) algorithm to learn discrete hash code.
    \item Column sampling based discrete supervised hashing~(COSDISH)~\cite{DBLP:conf/aaai/KangLZ16}: COSDISH is a supervised hashing method. COSDISH can directly learn discrete hash code.
    \item Deep hashing network~(DHN)~\cite{DBLP:conf/aaai/ZhuL0C16}: DHN is a deep supervised hashing method. DHN minimizes both pairwise cross-entropy loss and pairwise quantization loss.
    \item Deep supervised hashing~(DSH)~\cite{liudeep}: DSH is a deep supervised hashing method. DSH takes pairs of points as input and learns binary codes by maximizing the discriminability of the corresponding binary codes.
    \item Deep pairwise-supervised hashing~(DPSH)~\cite{DBLP:conf/ijcai/LiWK16}: DPSH is a deep supervised hashing method. DPSH performs simultaneous deep feature learning and hash-code learning with pairwise labels by minimizing negative log-likelihood of the observed pairwise labels.
\end{itemize}

Among all these baselines, LSH is a data-independent hashing method. ITQ is an unsupervised hashing method. LFH, FastH, COSDISH, and SDH are non-deep methods, which cannot perform deep feature learning. LFH is a relaxation-based method. FastH, COSDISH and SDH are discrete supervised hashing methods. DHN, DSH, and DPSH are deep hashing methods which can perform feature learning and hash-code learning simultaneously.

We first resize all images to be $224\times 224$ pixels for three datasets.
Then the raw image pixels are directly utilized as input for deep hashing methods.
For fair comparison, all deep hashing methods, including deep baselines and our DDSH, adopt the same pre-trained CNN-F model on ImageNet~\footnote{We download the CNN-F model pre-trained on ImageNet from~\url{http://www.vlfeat.org/matconvnet/pretrained/}.} for feature learning.
We carefully implement DHN and DSH on MatConvNet.
We fix the mini-batch size to be 128 and tune the learning rate from
$10^{-6}$ to $10^{-2}$ by using a cross-validation strategy.
Furthermore, we set weight decay as $5\times10^{-4}$ to avoid overfitting.
For DDSH, we set $\vert\Omega\vert=100$, $T_{out}=3$ and $T_{in}=50$.
Because NUS-WIDE is a multi-label dataset, we reduce the similarity weight for those training points with multi-labels when we train DDSH.

For non-deep hashing methods, including LFH, ITQ, LFH, FastH, SDH and COSDISH, we use 4,096-dim deep features extracted by the CNN-F model pre-trained on ImageNet as input for fair comparison.
Because SDH is a kernel-based methods,
we randomly sample 1,000 data points as anchors to construct the kernel by following the suggestion of the authors of SDH~\cite{DBLP:conf/cvpr/ShenSLS15}.
For LFH, FastH and COSDISH,
we utilize boosted decision tree for out-of-sample extension by following the setting of FastH.

In our experiment, ground-truth neighbors are defined based on whether two data points share at least one class label. We carry out Hamming ranking task and hash lookup task to evaluate DDSH and baselines. We report the Mean Average Precision~(MAP), Top-K precision, precision-recall curve and case study for Hamming ranking task. Specifically, given a query $\x_q$, we can calculate its average precision~(AP) through the following equation:
\begin{align}
AP(\x_q)=\frac{1}{R_k}\sum_{k=1}^N P(k)\IB_1(k), \nonumber
\end{align}
where $R_k$ is the number of the relevant samples, $P(k)$ is the precision at cut-off $k$ in the returned sample list and $\IB_1(k)$ is an indicator function which equals 1 if the $k$th returned sample is a ground-truth  neighbor of $\x_q$. Otherwise, $\IB_1(k)$ is 0. Given $Q$ queries, we can compute the MAP as follows:
\begin{align}
MAP=\frac{1}{Q}\sum_{q=1}^QAP(\x_q). \nonumber
\end{align}
Because NUS-WIDE is relatively large, the MAP value on NUS-WIDE is calculated based on the top 5000 returned neighbors. The MAP values for other datasets are calculated based on the whole retrieval set.

For hash lookup task, we report mean hash lookup success rate~(SR) within Hamming radius 0, 1 and 2~\cite{DBLP:conf/nips/LiuMKC14}.
When at least one ground-truth neighbor is retrieved within a specific Hamming radius, we call it a lookup success. The hash lookup success rate~(SR) can be calculated as follows:
\begin{align}\small
SR=\sum_{q=1}^Q\frac{\IB(\text{number of retrieved ground-truth for query }{\x}_q > 0)}{Q}\nonumber
\end{align}
Here, $\IB(\cdot)$ is an indicator function, i.e., $\IB(\text{true})=1$ and $\IB(\text{false})=0$.
$Q$ is the total number of query images.  All experiments are run 5 times, and the average performance is reported.

\subsection{Experimental Result}

\subsubsection{Hamming Ranking Task}

Table~\ref{tab:map} reports the MAP result on three datasets. We can easily find that our DDSH achieves the state-of-the-art retrieval accuracy in all cases compared with all baselines, including deep hashing methods, non-deep supervised hashing methods, non-deep unsupervised hashing methods and data-independent methods.

\begin{table*}[!htb]
\centering
\caption{MAP of the Hamming ranking task. The best accuracy is shown in boldface.}
\label{tab:map}
\begin{tabular}{|c||c|c|c|c||c|c|c|c||c|c|c|c|}
 \hline
 \multirow{2}{*}{Method} &
 \multicolumn{4}{c||}{{CIFAR-10}} & \multicolumn{4}{c||}{{SVHN}}& \multicolumn{4}{c|}{{NUS-WIDE}}\\
 \cline{2-13} & 12 bits & 24 bits & 32 bits & 48 bits  & 12 bits & 24 bits & 32 bits & 48 bits & 12 bits & 24 bits & 32 bits & 48 bits \\
 \hline \hline
DDSH & {\bf 0.769} & {\bf 0.829} & {\bf 0.835} & {\bf 0.819} & {\bf 0.574} & {\bf 0.674} & {\bf 0.703} & {\bf 0.718}
& {\bf 0.791} & {\bf 0.815} & {\bf 0.821} & {\bf 0.827}\\
\shline
DSH & 0.646 & {0.749} & {0.786} & {0.811} & 0.370 & {0.480} & {0.523} & {0.583} & 0.762 & 0.794 & 0.797 & 0.808\\
\hline
DPSH & {0.684} & 0.723 & 0.740 & 0.746 & 0.379 & 0.422 & 0.434 & 0.456& 0.788 & 0.809 & {0.817} & {0.823}\\
\hline
DHN & 0.673 & 0.711 & 0.705 & 0.713 & {0.380} & 0.410 & 0.416 & 0.430  & {0.790} & {0.810} & 0.809 & 0.818\\
\shline
COSDISH  & 0.609 & 0.683 & 0.696 & 0.716  & 0.238 & 0.295 & 0.320 & 0.341  &  0.730 & 0.764 & 0.787 & 0.799  \\
\hline
SDH & 0.520 & 0.646 & 0.658 & 0.669 & 0.151 & 0.300 & 0.320 & 0.334 & 0.739 & 0.762 & 0.770 & 0.772 \\
\hline
FastH & 0.620 & 0.673 & 0.687 & 0.716 & 0.252 & 0.296 & 0.318 & 0.344  & 0.741 & 0.783 & 0.795 & 0.809 \\
\hline
LFH & 0.401 & 0.605 & 0.657 & 0.700 & 0.193 & 0.256 & 0.284 & 0.325 & 0.705 & 0.759 & 0.778 & 0.794 \\
\shline
ITQ & 0.258 & 0.272 & 0.283 & 0.294 & 0.111 & 0.114 & 0.115 & 0.116 & 0.505 & 0.504 & 0.503 & 0.505 \\
\hline
LSH & 0.147 & 0.172 & 0.180 & 0.193 & 0.107 & 0.108 & 0.109 & 0.111 & 0.341 & 0.351 & 0.351 & 0.371 \\
\hline
 \end{tabular}
\end{table*}

\begin{figure*}[htb]
\centering
\small
\begin{tabular}{c@{ }@{ }c@{ }@{ }c@{ }@{ }c}
\begin{minipage}{0.24\linewidth}\centering
    \includegraphics[width=1\textwidth]{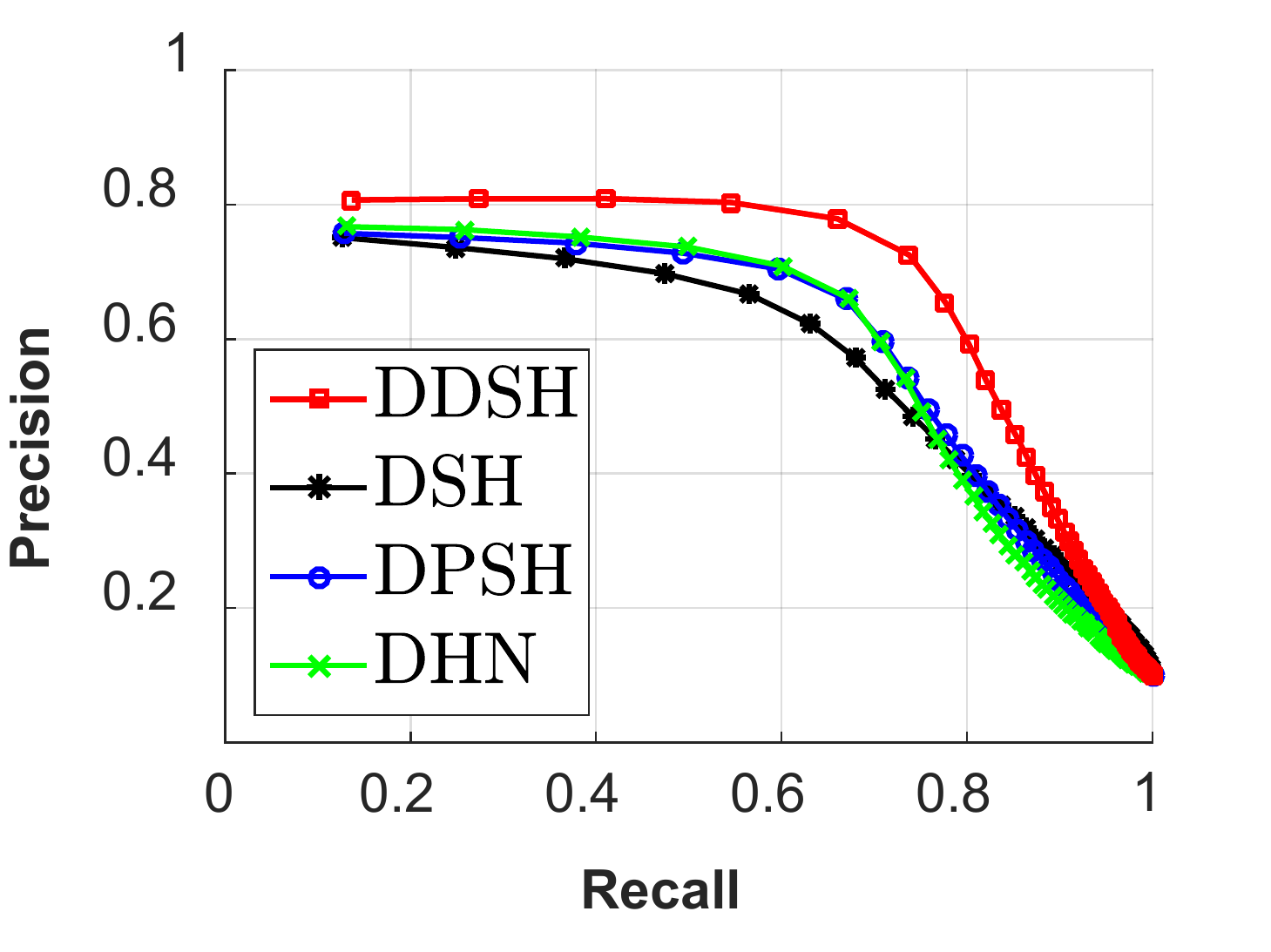}\\
    (a) 12 bits @CIFAR-10
\end{minipage} &
\begin{minipage}{0.24\linewidth}\centering
    \includegraphics[width=1\textwidth]{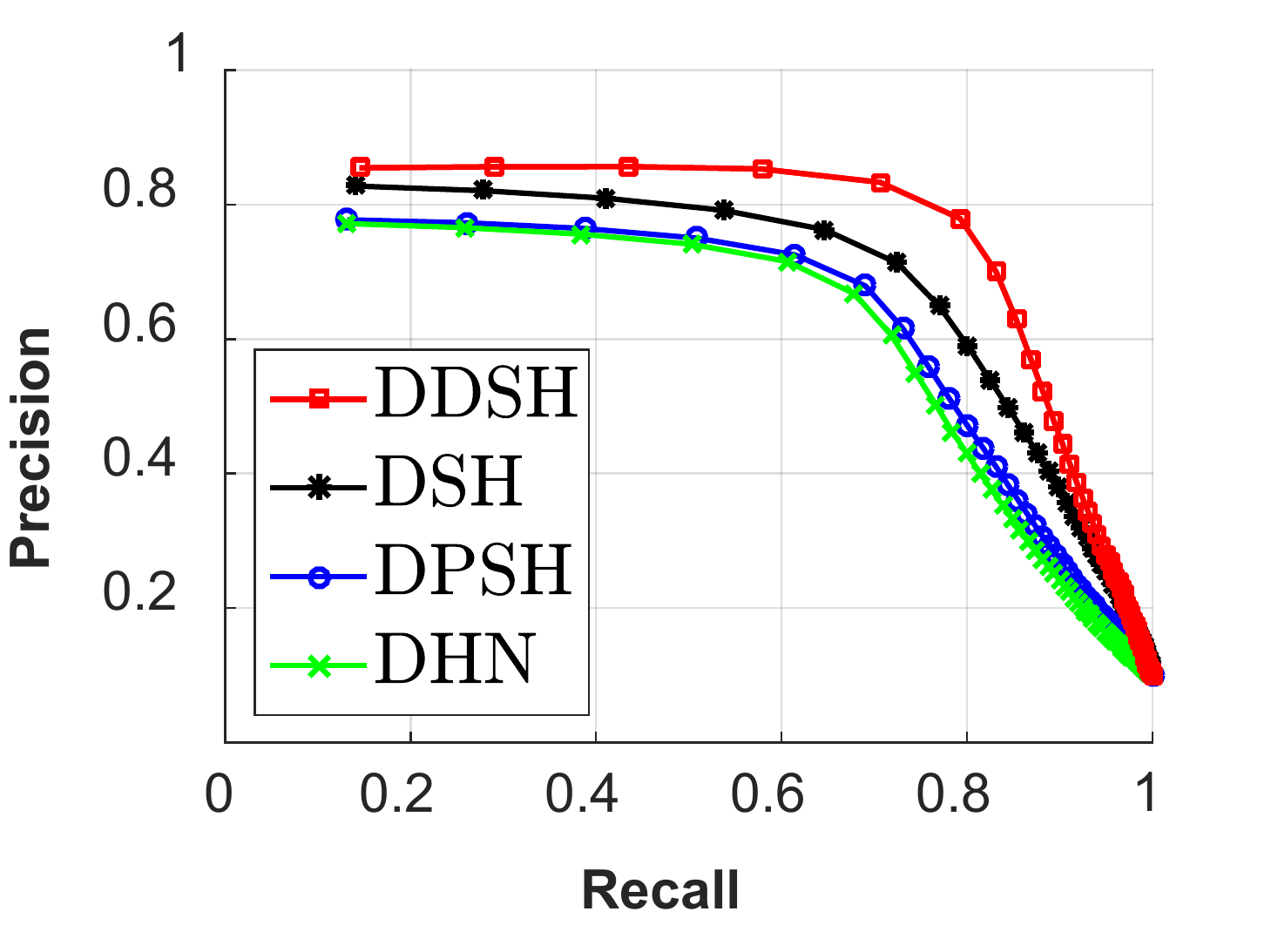}\\
    (b) 24 bits @CIFAR-10
\end{minipage} &
\begin{minipage}{0.24\linewidth}\centering
    \includegraphics[width=1\textwidth]{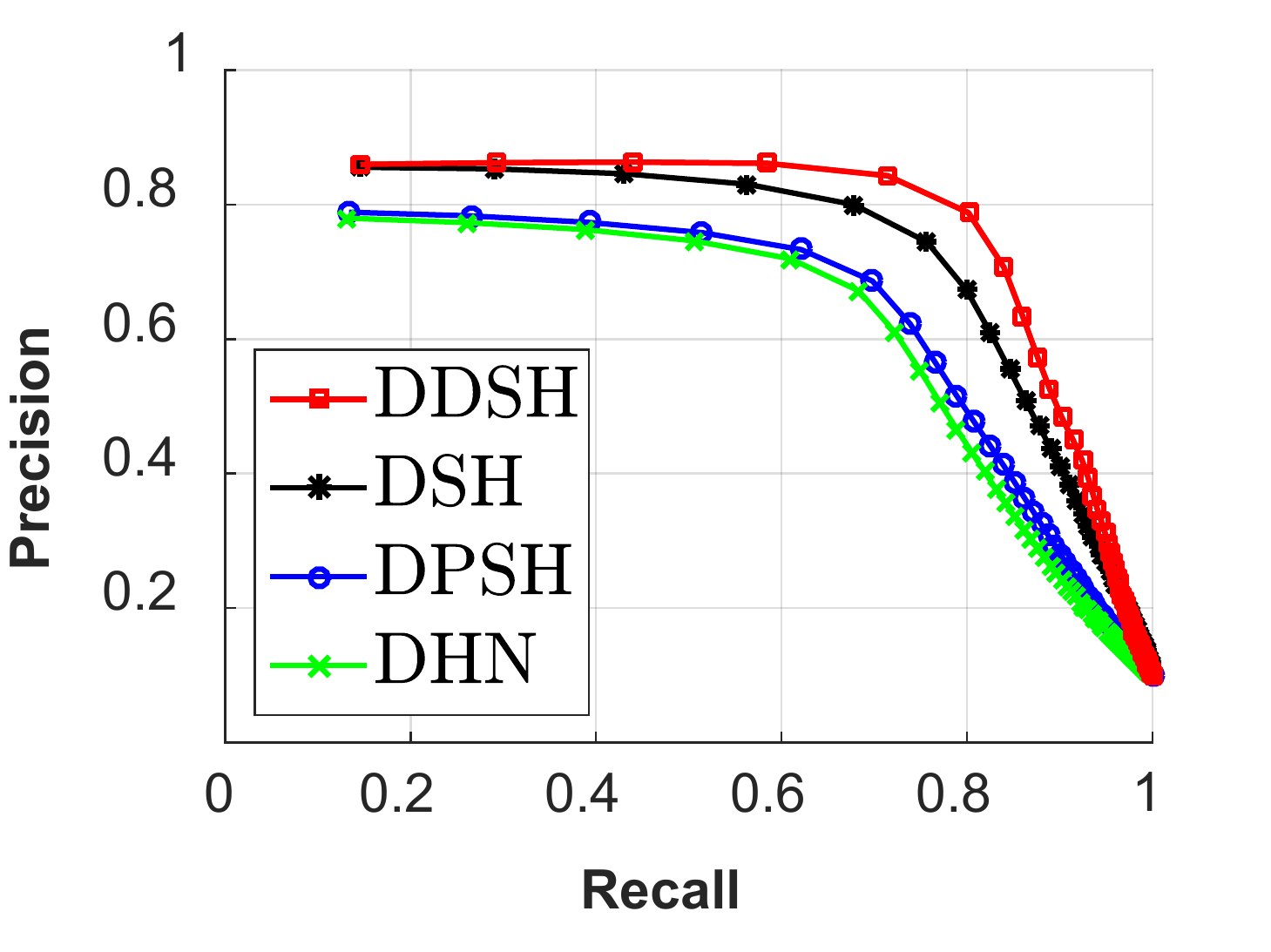}\\
    (c) 32 bits @CIFAR-10
\end{minipage} &
\begin{minipage}{0.24\linewidth}\centering
    \includegraphics[width=1\textwidth]{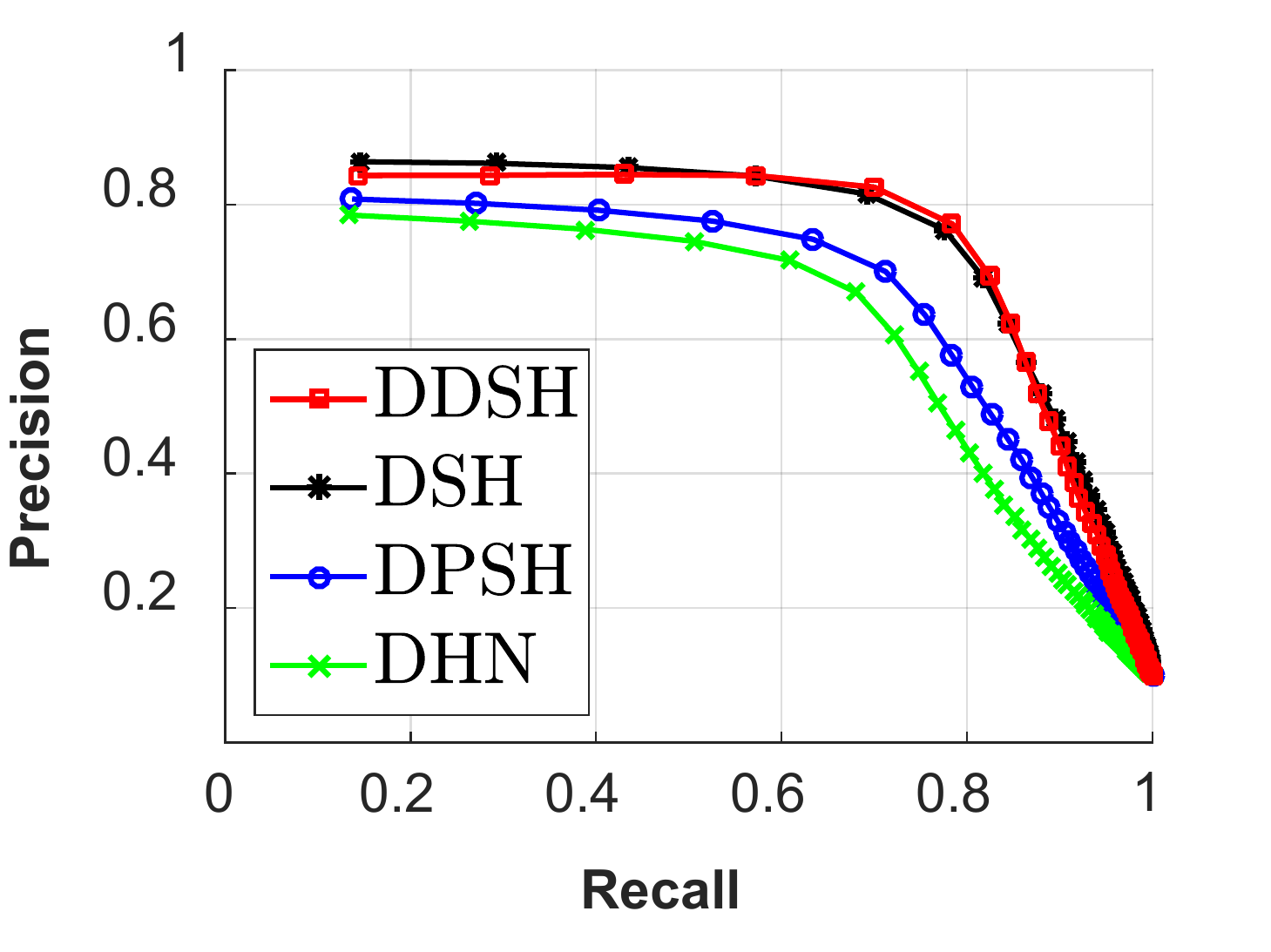}\\
    (d) 48 bits @CIFAR-10
\end{minipage} \vspace{5pt}\\
\begin{minipage}{0.24\linewidth}\centering
    \includegraphics[width=1\textwidth]{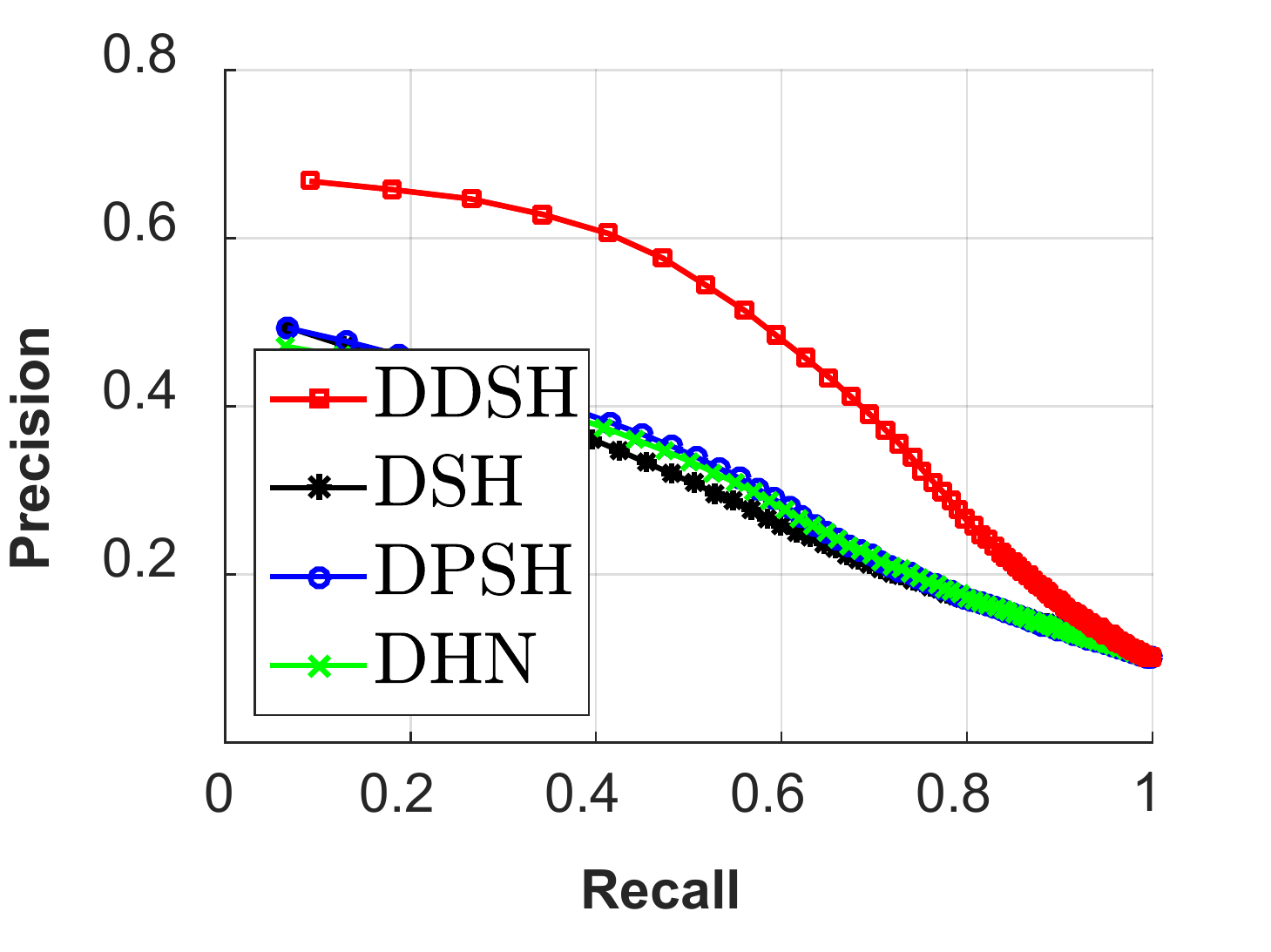}\\
    (e) 12 bits @SVHN
\end{minipage} &
\begin{minipage}{0.24\linewidth}\centering
    \includegraphics[width=1\textwidth]{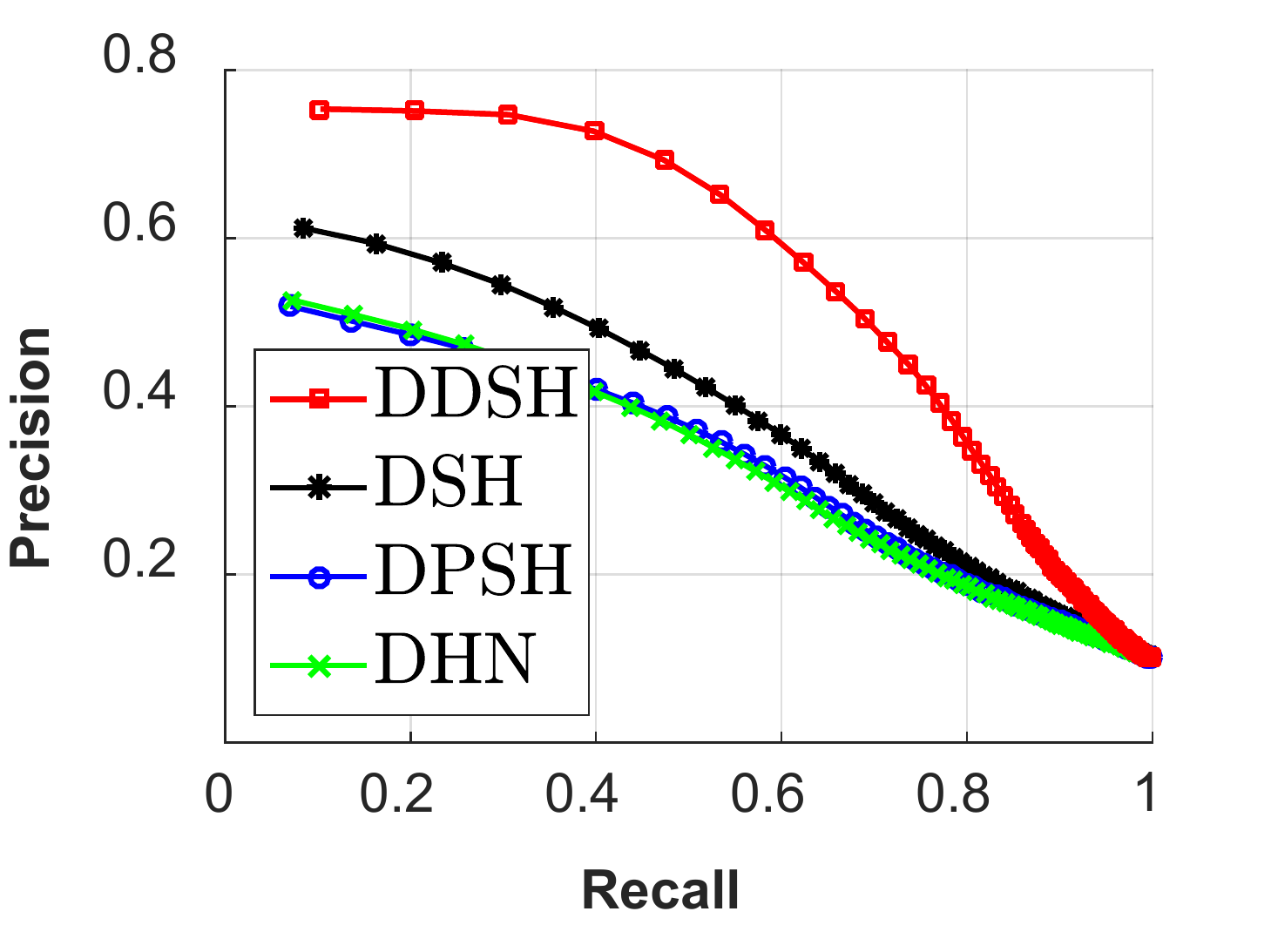}\\
    (f) 24 bits @SVHN
\end{minipage} &
\begin{minipage}{0.24\linewidth}\centering
    \includegraphics[width=1\textwidth]{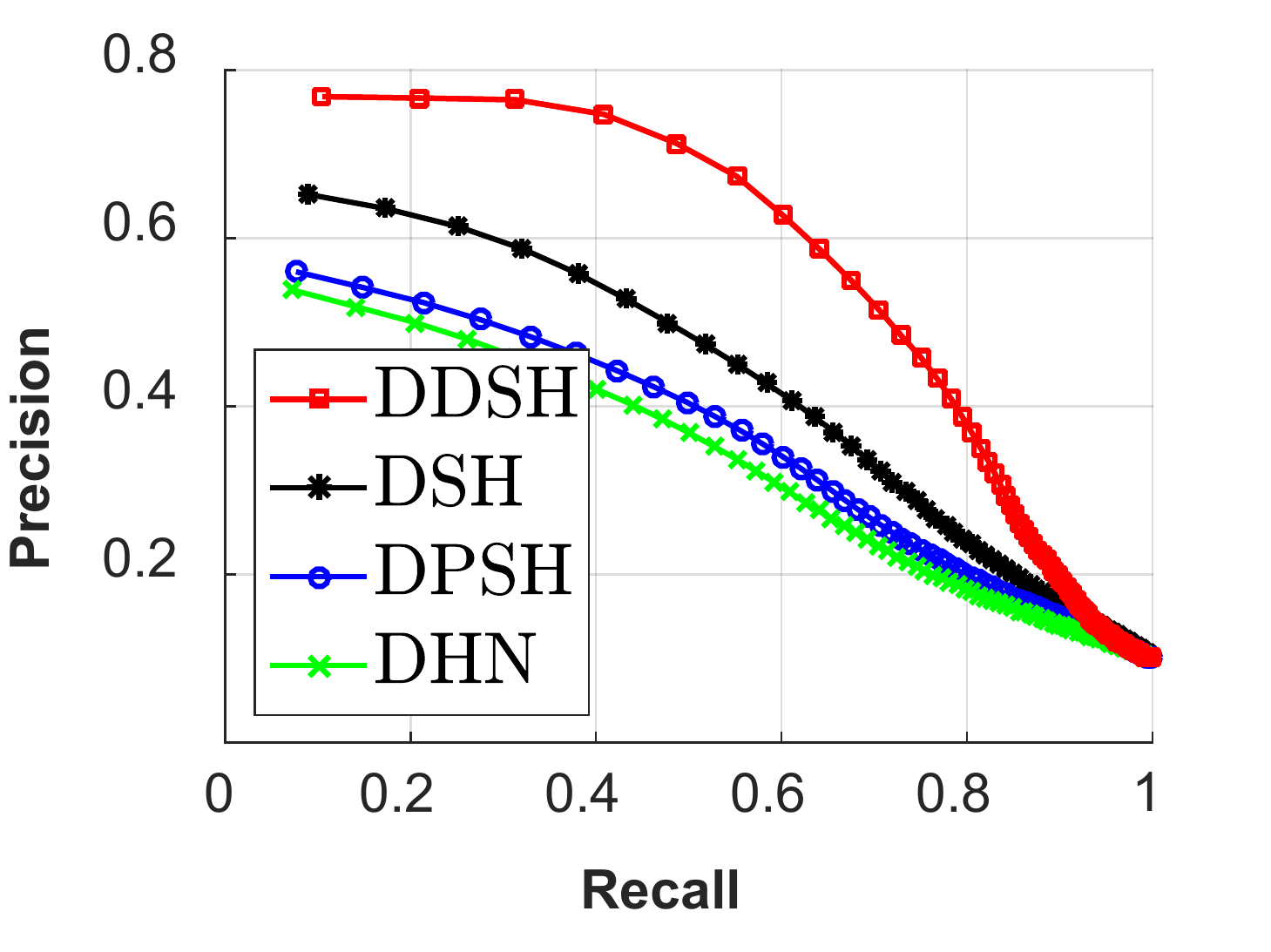}\\
    (g) 32 bits @SVHN
\end{minipage} &
\begin{minipage}{0.24\linewidth}\centering
    \includegraphics[width=1\textwidth]{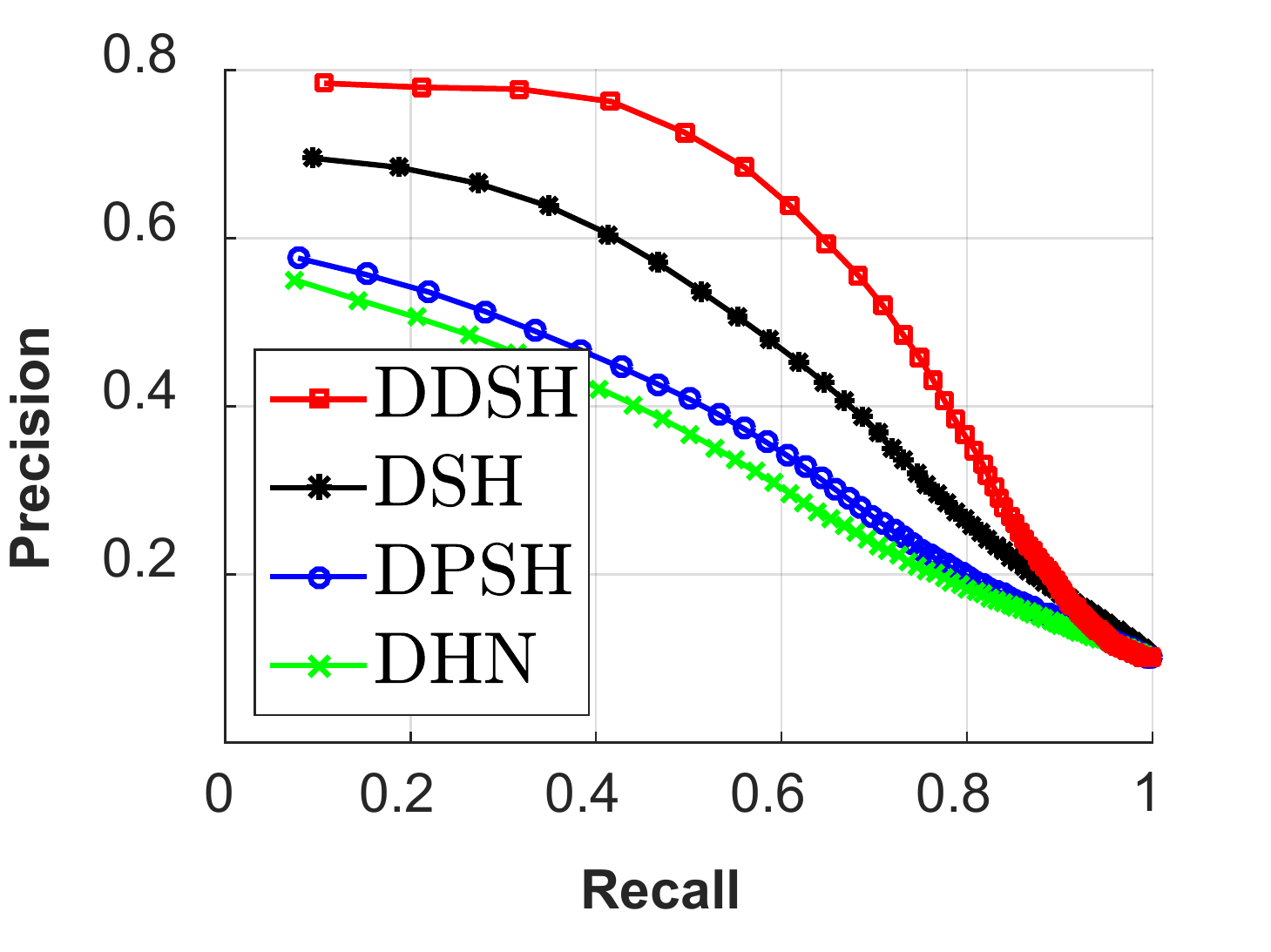}\\
    (h) 48 bits @SVHN
\end{minipage} \vspace{5pt}\\
\begin{minipage}{0.24\linewidth}\centering
    \includegraphics[width=1\textwidth]{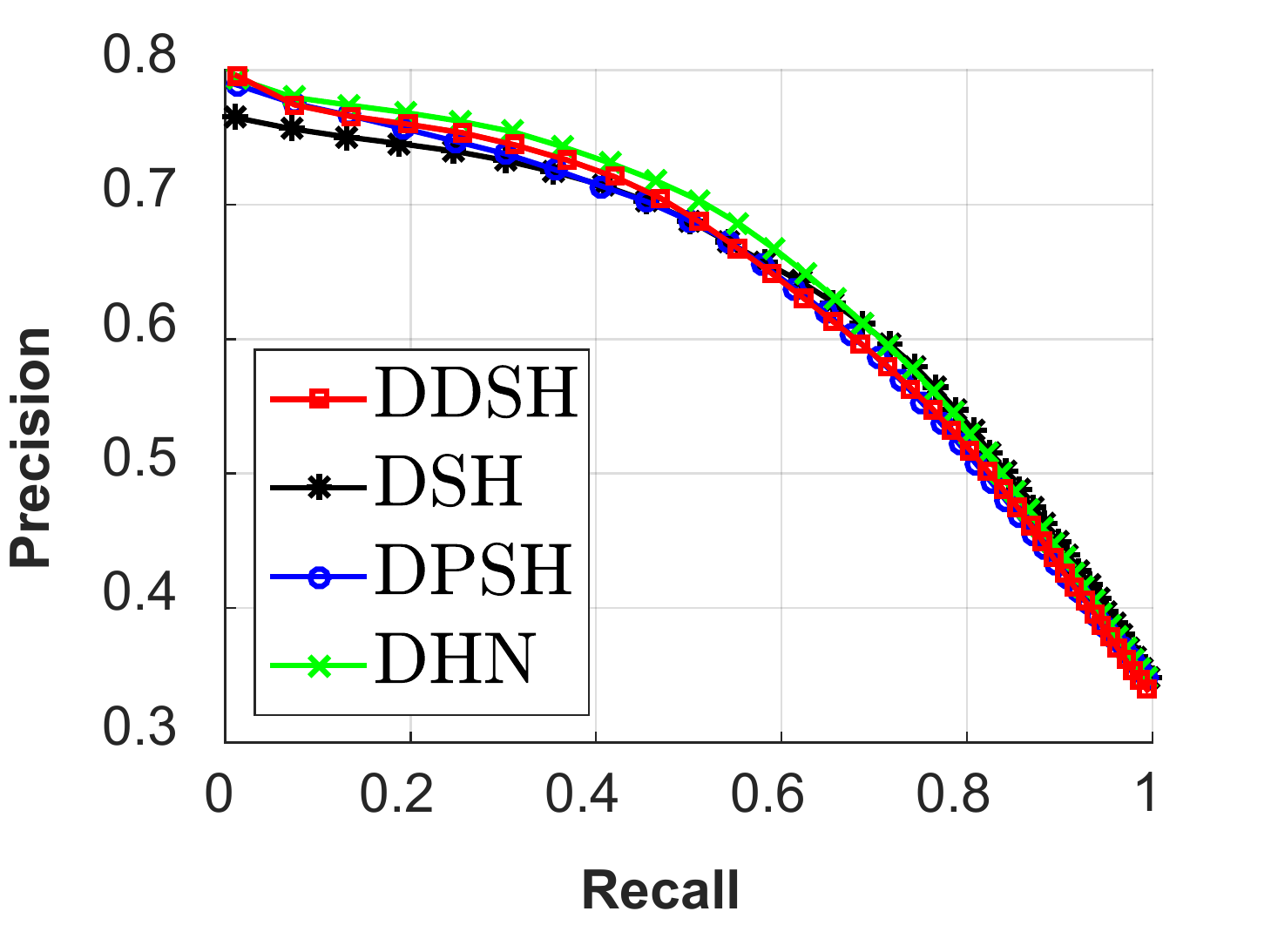}\\
    (i) 12 bits @NUS-WIDE
\end{minipage} &
\begin{minipage}{0.24\linewidth}\centering
    \includegraphics[width=1\textwidth]{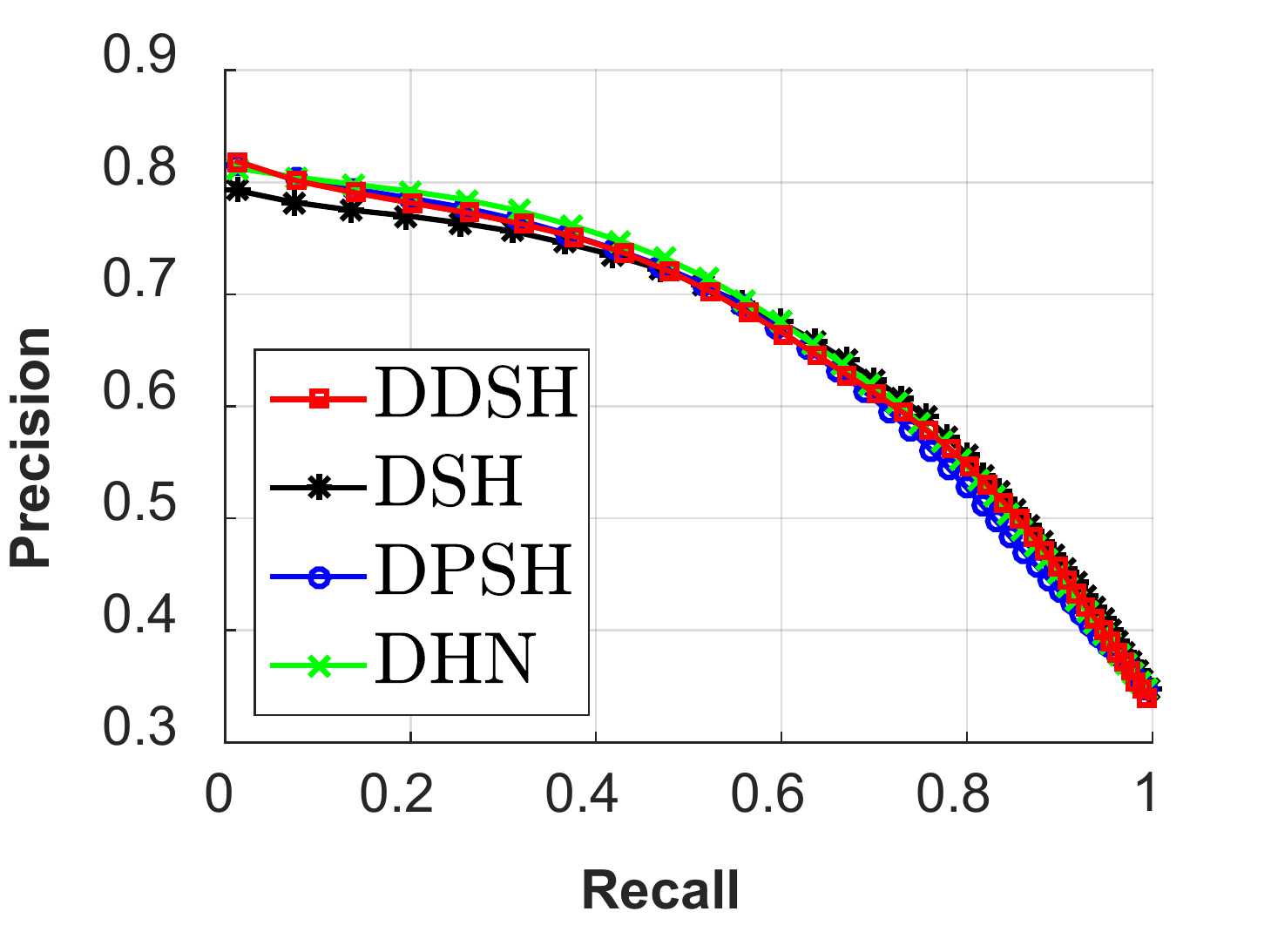}\\
    (j) 24 bits @NUS-WIDE
\end{minipage} &
\begin{minipage}{0.24\linewidth}\centering
    \includegraphics[width=1\textwidth]{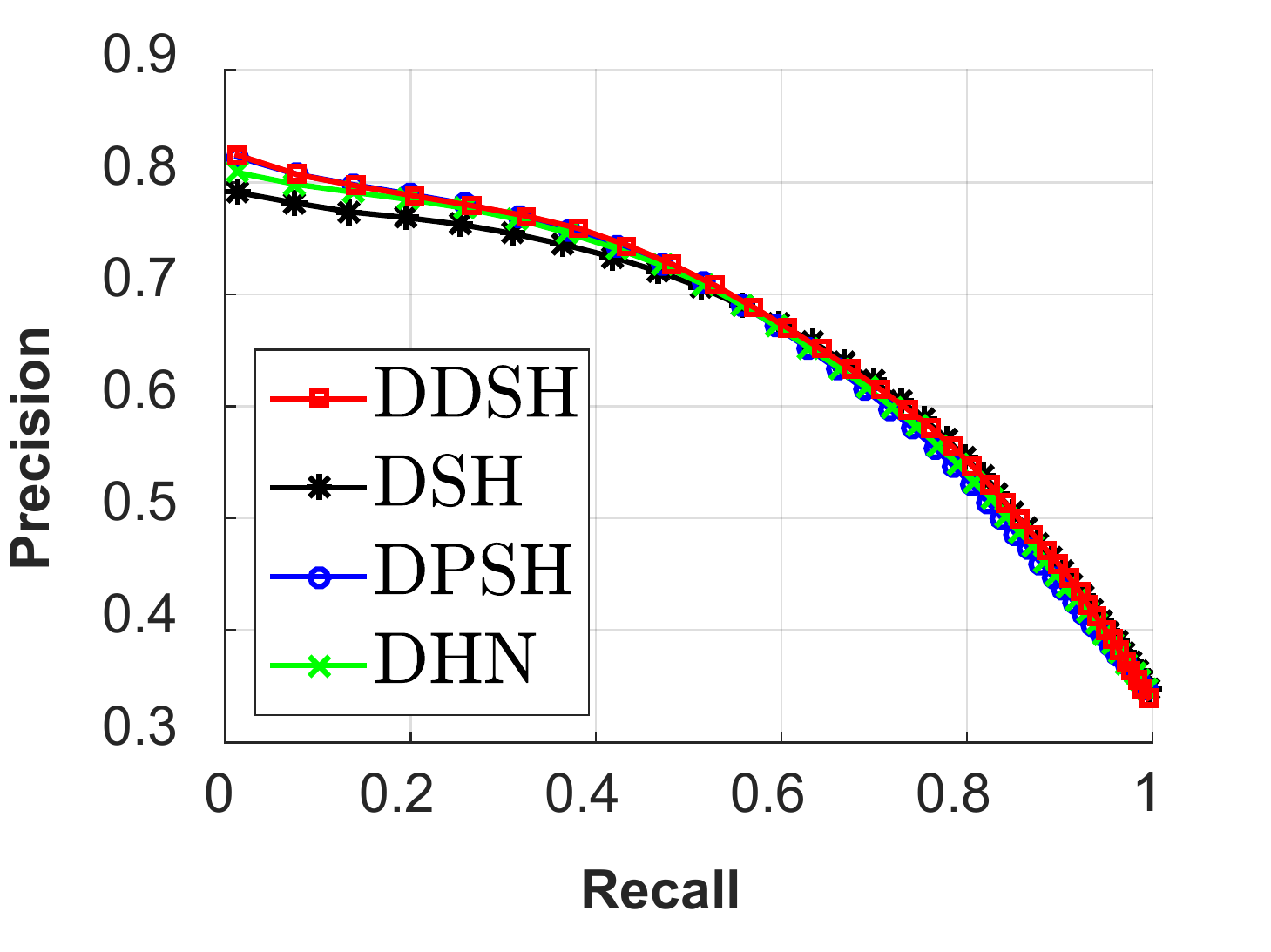}\\
    (k) 32 bits @NUS-WIDE
\end{minipage} &
\begin{minipage}{0.24\linewidth}\centering
    \includegraphics[width=1\textwidth]{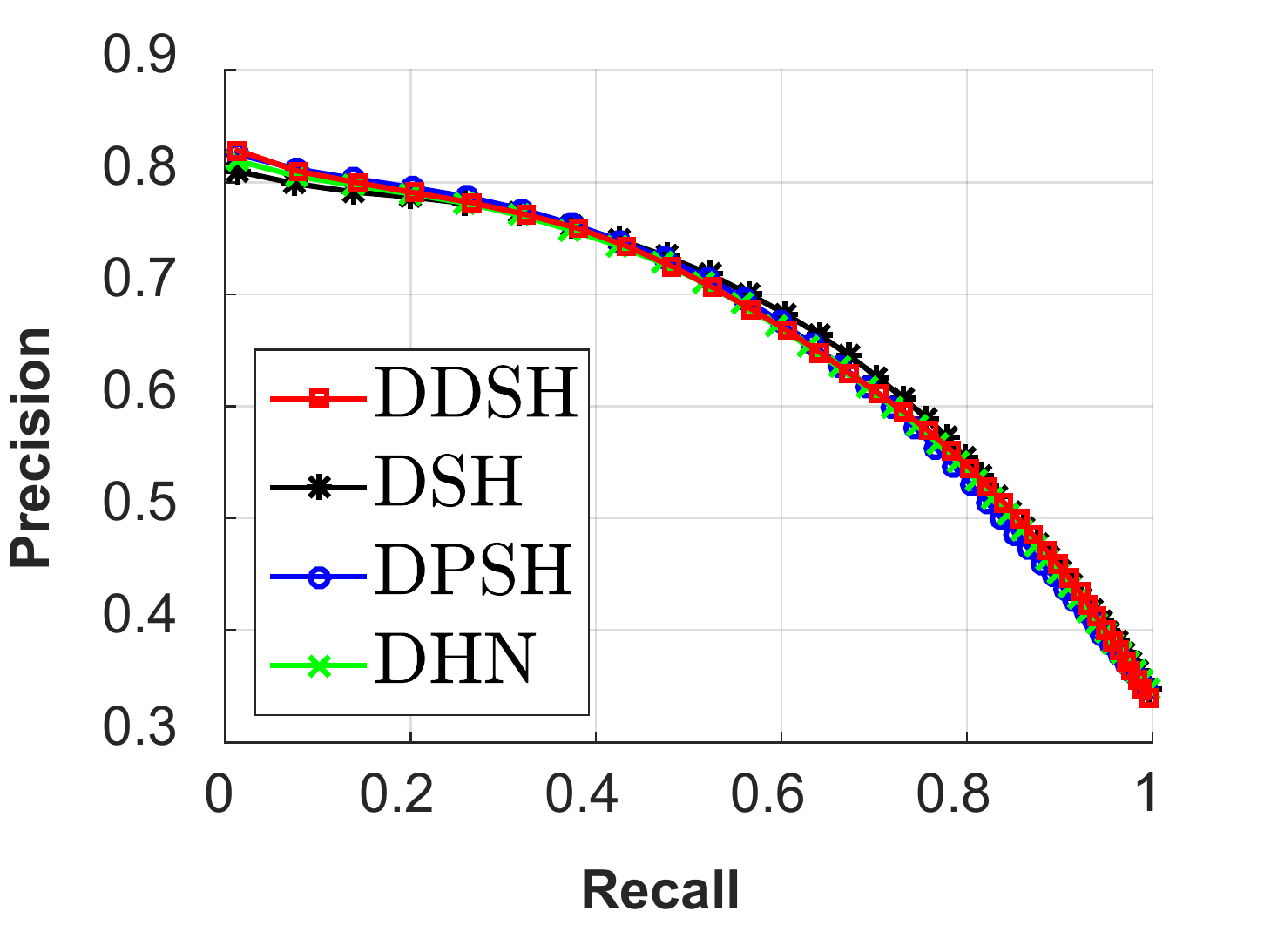}\\
    (l) 48 bits @NUS-WIDE
\end{minipage}
\end{tabular}\vspace*{0pt}
\caption{Performance of precision-recall curve on three datasets. The four sub-figures in each row are the precision-recall curves for 12 bits, 24 bits, 32 bits and 48 bits respectively.}
\label{fig:pr}
\end{figure*}

By comparing ITQ to LSH, we can find that the data-dependent hashing methods can significantly outperform data-independent hashing methods. By comparing COSDISH, SDH, FastH and LFH to ITQ, we can find that supervised methods can outperform unsupervised methods because of the effect of using supervised information. By comparing COSDISH, SDH and FastH to LFH, we can find that discrete supervised hashing can outperform relaxation-based continuous hashing, which means that discrete coding procedure is able to learn more optimal binary codes. By comparing deep hashing methods, i.e., DDSH, DPSH, DHN and DSH, to non-deep hashing methods, we can find that deep hashing can outperform non-deep hashing because deep supervised hashing can perform deep feature learning compared with non-deep hashing methods. This experimental result demonstrates that deep supervised hashing is a more compatible architecture for hashing learning.

The main difference between our proposed DDSH and other discrete supervised hashing methods like COSDISH, SDH and FastH is that our DDSH adopts supervised information to directly guide deep feature learning procedure but other discrete supervised hashing methods do not have deep feature learning ability. The main difference between our DDSH and other deep hashing methods is that DDSH adopts supervised information to directly guide the discrete coding procedure but other deep hashing methods do not have this property. Hence, the experimental results successfully demonstrate the motivation of DDSH, i.e., utilizing supervised information to \emph{directly} guide both deep feature learning procedure and discrete coding procedure can further improve retrieval performance in real applications.

Furthermore, we select three best baselines, i.e., DSH, DPSH and DHN, to compare the precision-recall and top-k precision results. We report the precision-recall curve on all three datasets in Figure~\ref{fig:pr}. We can see that the proposed DDSH still achieves the best performance in terms of precision-recall curve in most cases.

\begin{figure*}[htb]
\centering
\small
\begin{tabular}{c@{ }@{ }c@{ }@{ }c@{ }@{ }c}
\begin{minipage}{0.24\linewidth}\centering
    \includegraphics[width=1\textwidth]{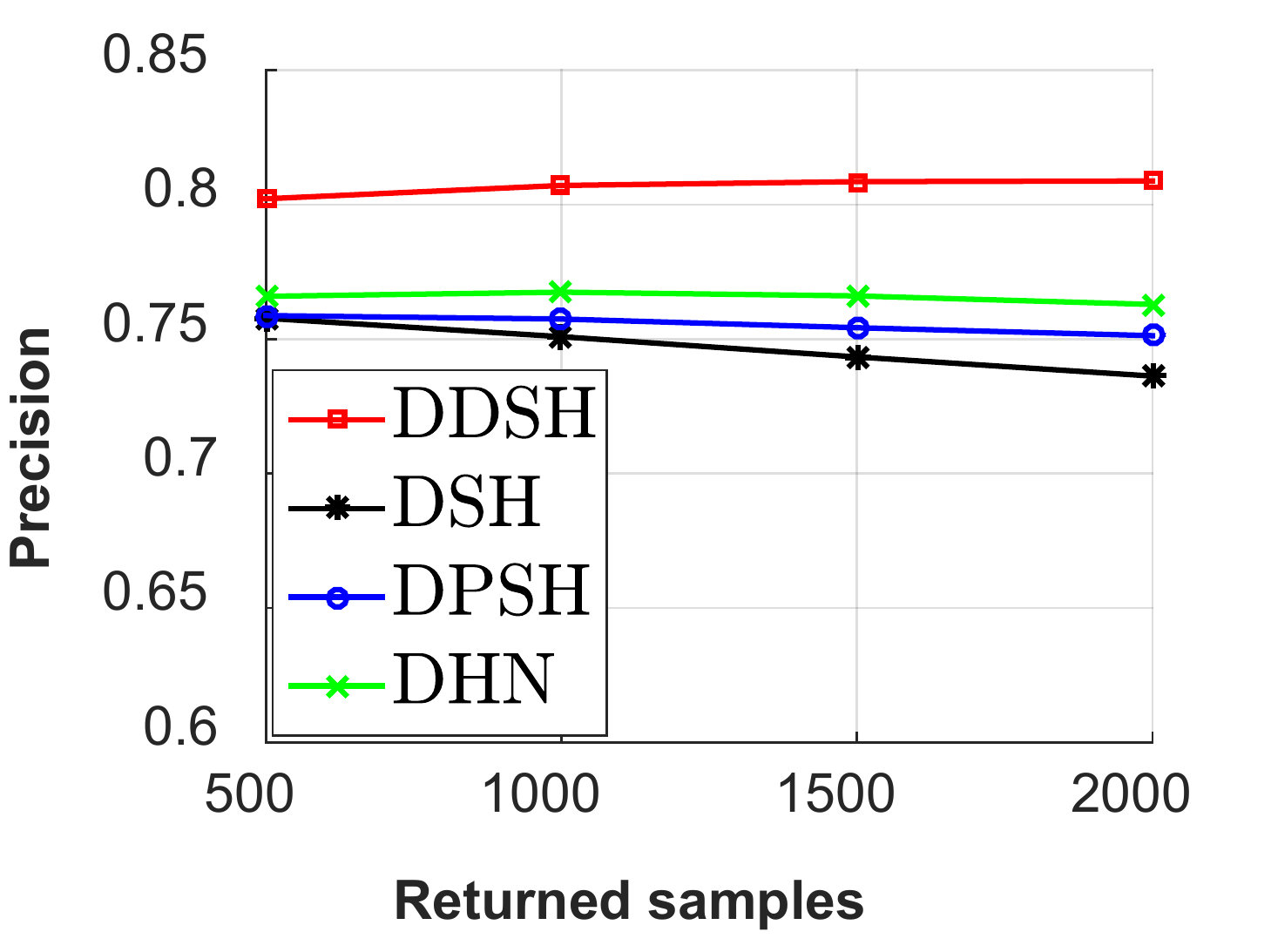}\\
    (a) 12 bits @CIFAR-10
\end{minipage} &
\begin{minipage}{0.24\linewidth}\centering
    \includegraphics[width=1\textwidth]{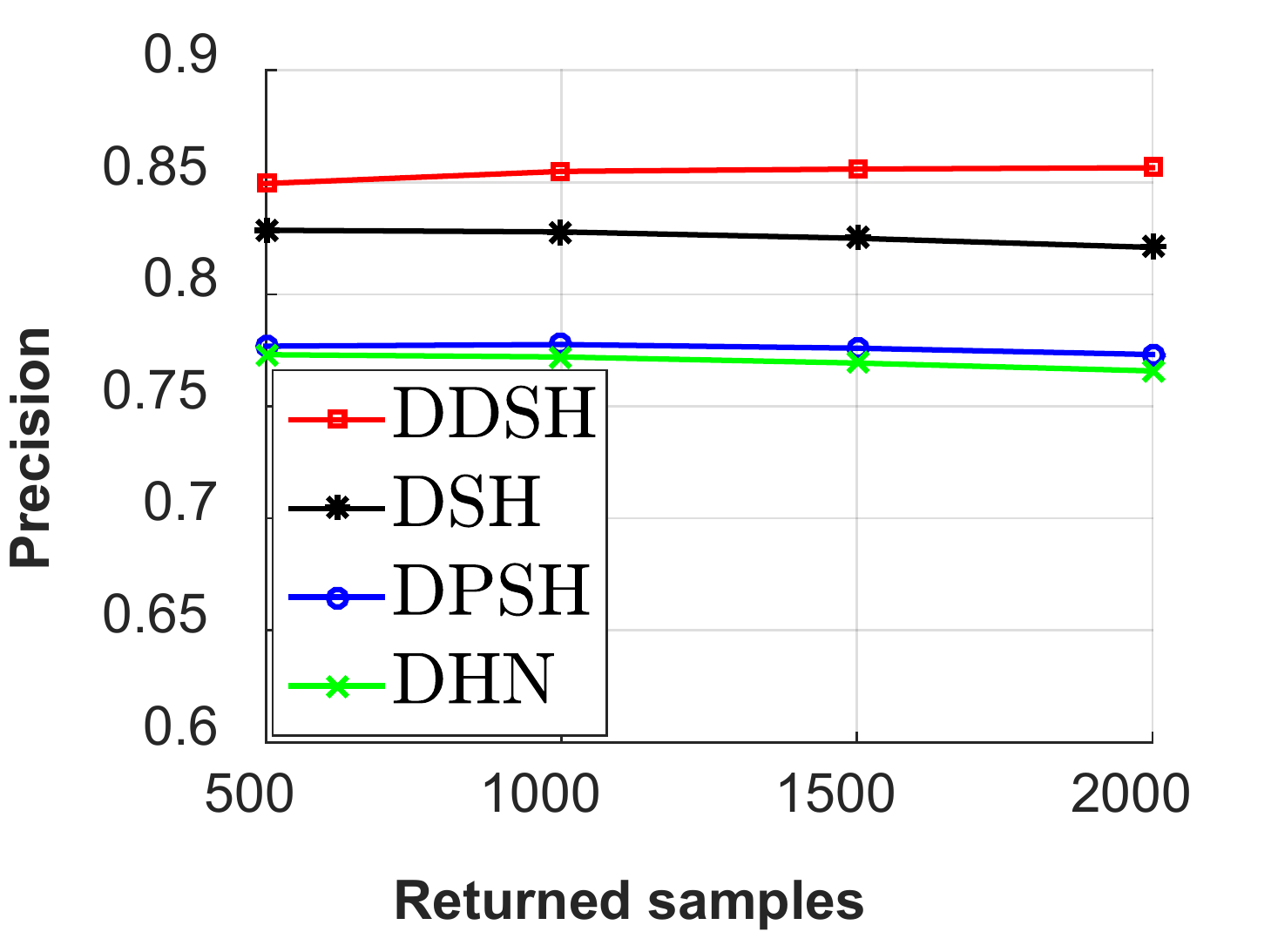}\\
    (b) 24 bits @CIFAR-10
\end{minipage} &
\begin{minipage}{0.24\linewidth}\centering
    \includegraphics[width=1\textwidth]{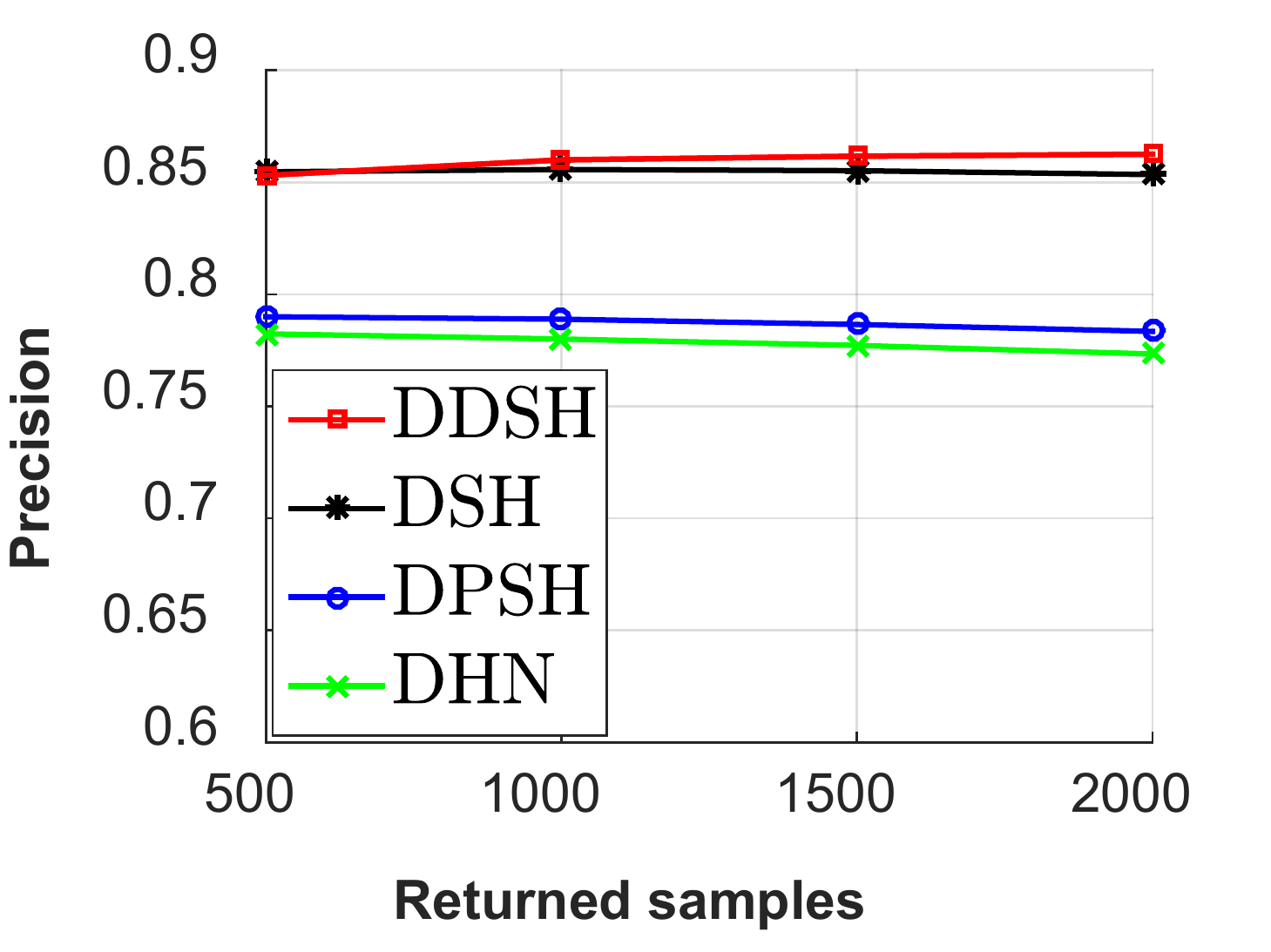}\\
    (c) 32 bits @CIFAR-10
\end{minipage} &
\begin{minipage}{0.24\linewidth}\centering
    \includegraphics[width=1\textwidth]{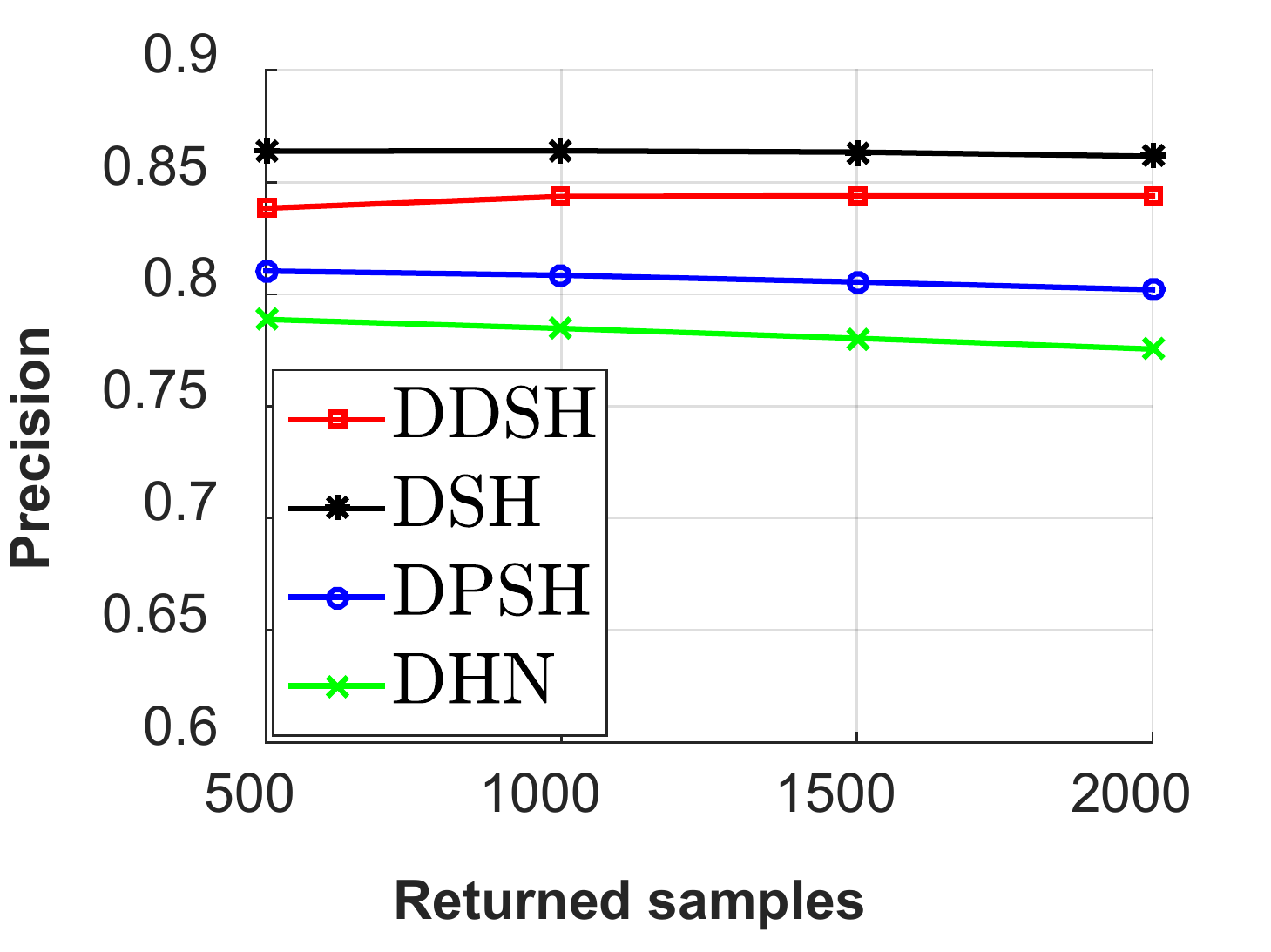}\\
    (d) 48 bits @CIFAR-10
\end{minipage} \vspace{5pt}\\
\begin{minipage}{0.24\linewidth}\centering
    \includegraphics[width=1\textwidth]{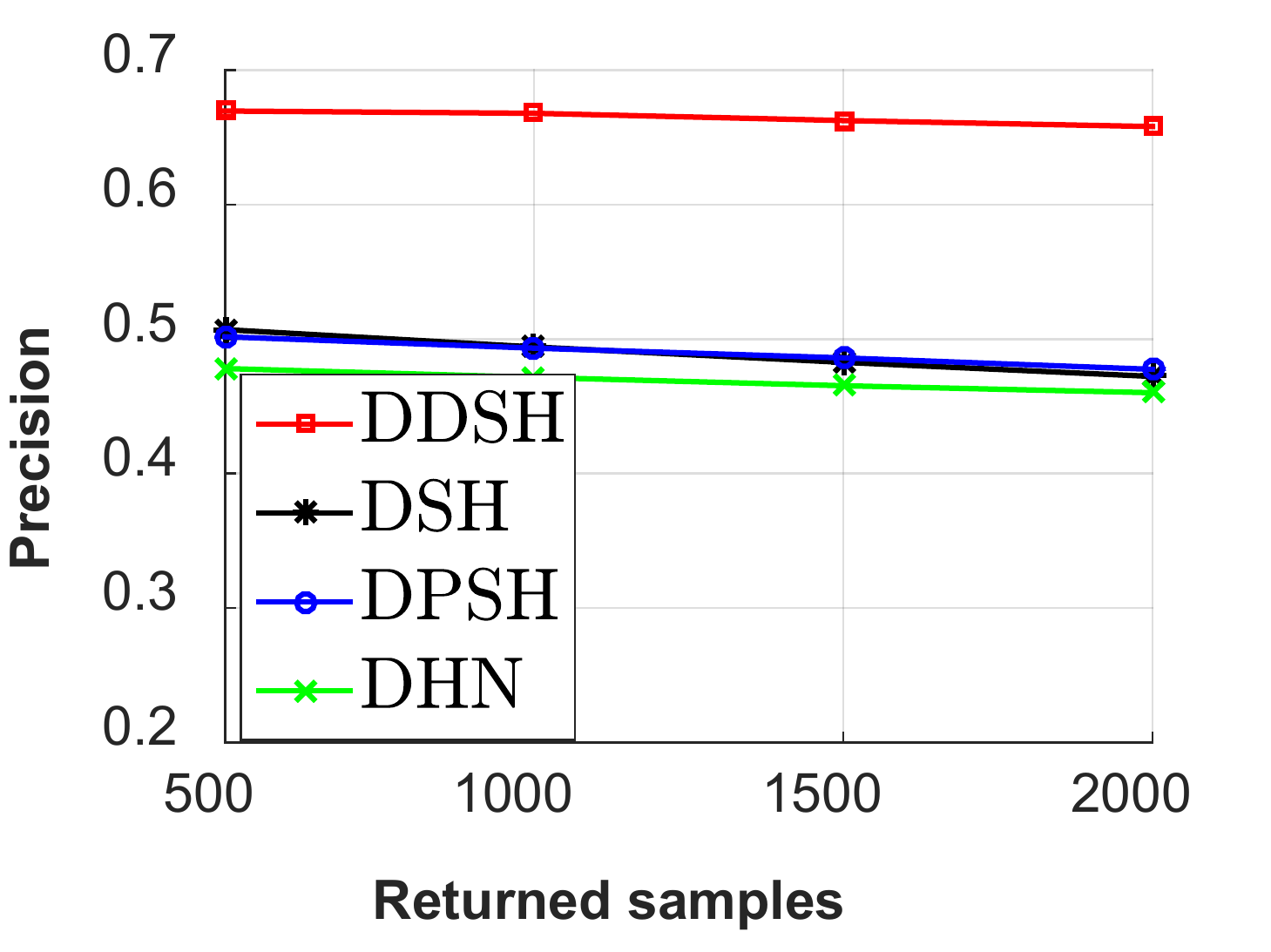}\\
    (e) 12 bits @SVHN
\end{minipage} &
\begin{minipage}{0.24\linewidth}\centering
    \includegraphics[width=1\textwidth]{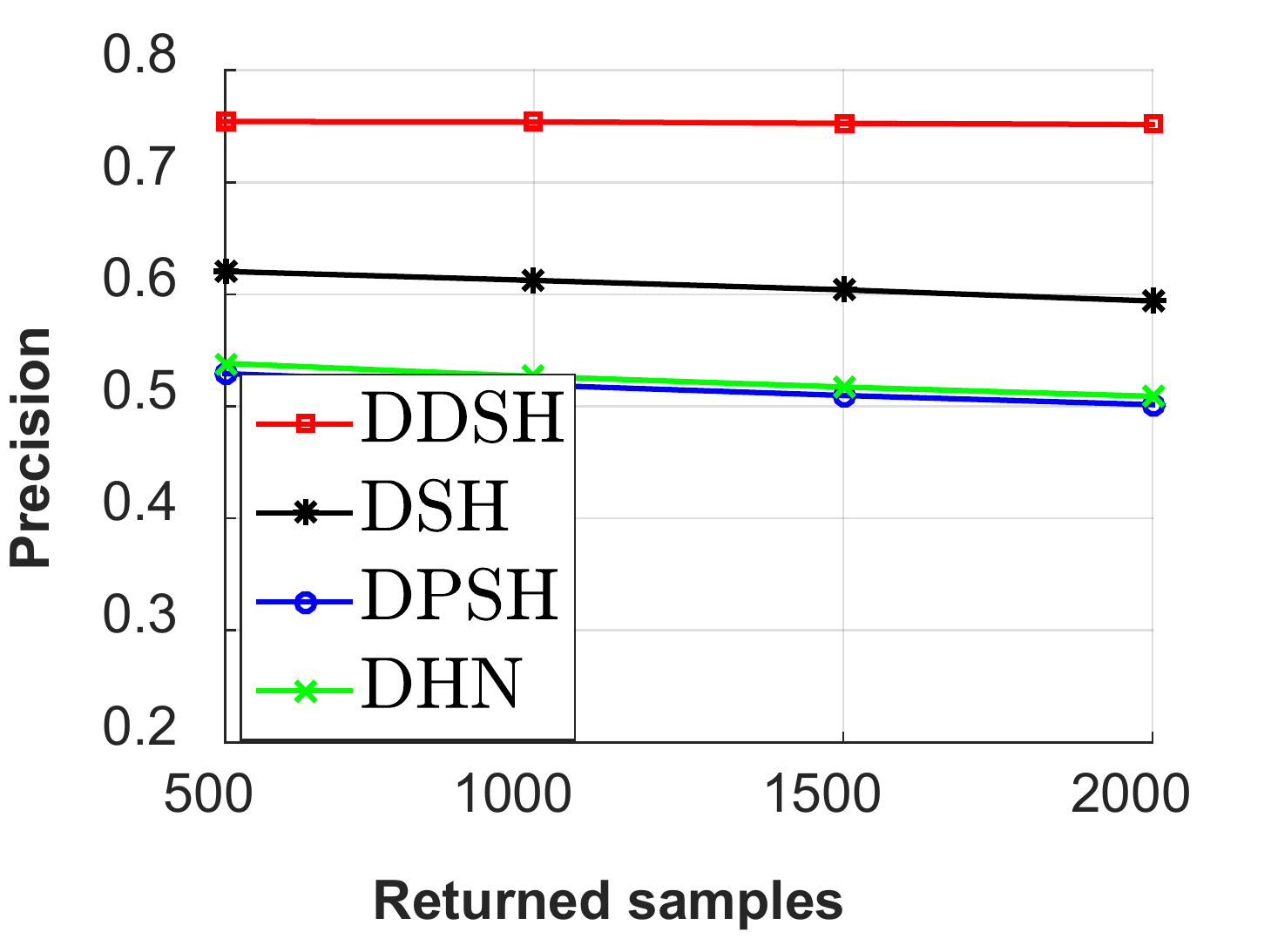}\\
    (f) 24 bits @SVHN
\end{minipage} &
\begin{minipage}{0.24\linewidth}\centering
    \includegraphics[width=1\textwidth]{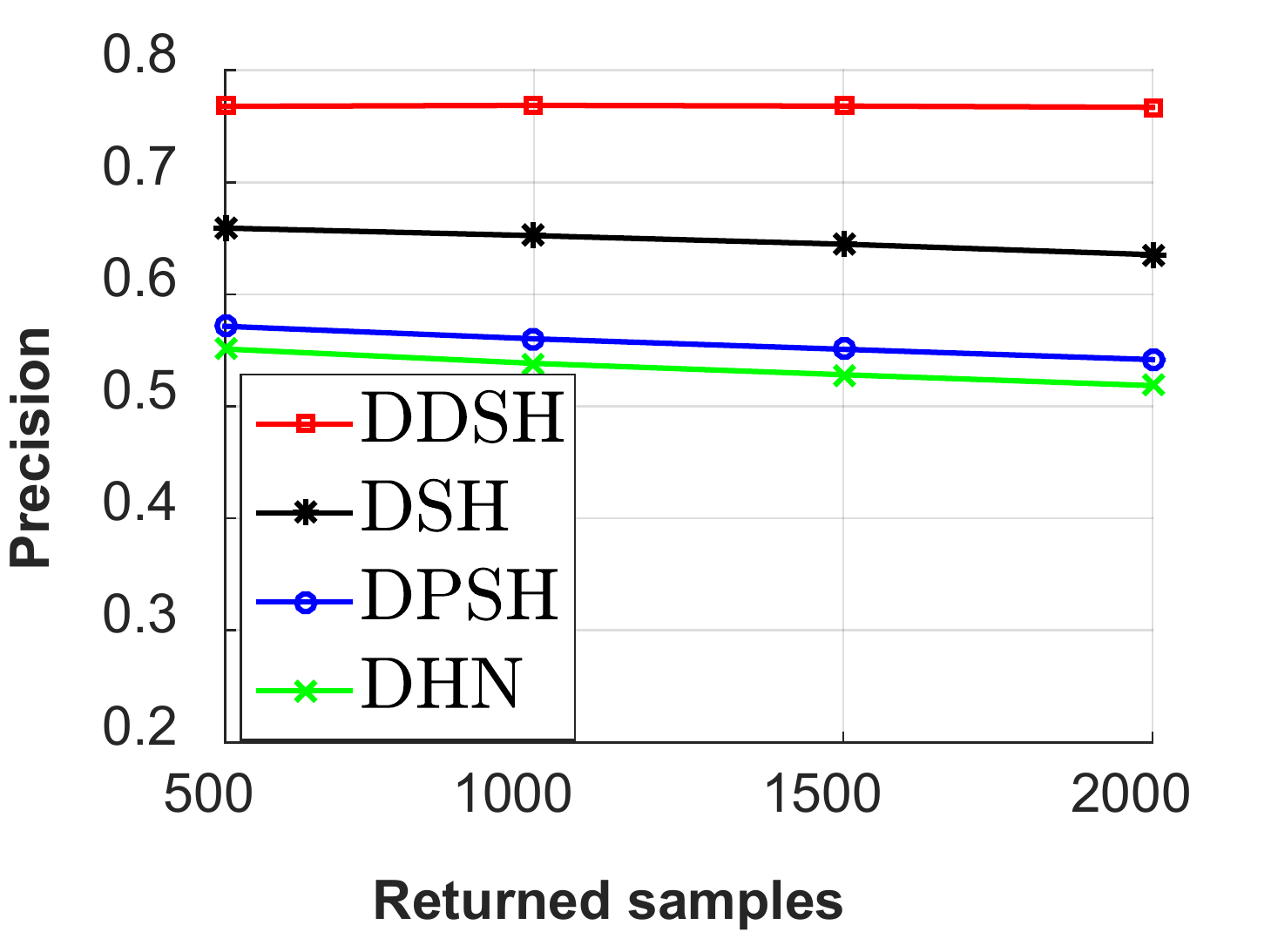}\\
    (g) 32 bits @SVHN
\end{minipage} &
\begin{minipage}{0.24\linewidth}\centering
    \includegraphics[width=1\textwidth]{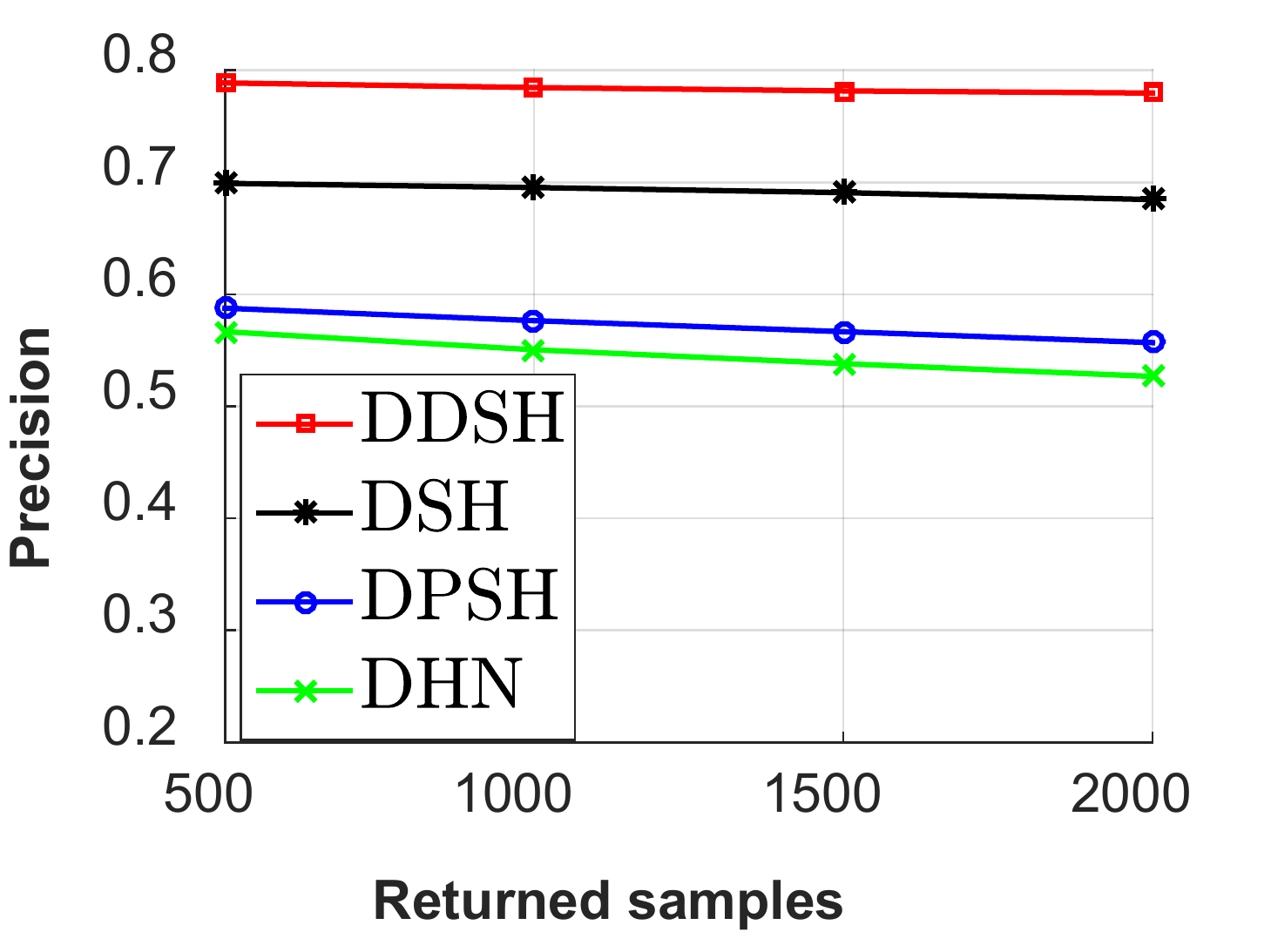}\\
    (h) 48 bits @SVHN
\end{minipage} \vspace{5pt}\\
\begin{minipage}{0.24\linewidth}\centering
    \includegraphics[width=1\textwidth]{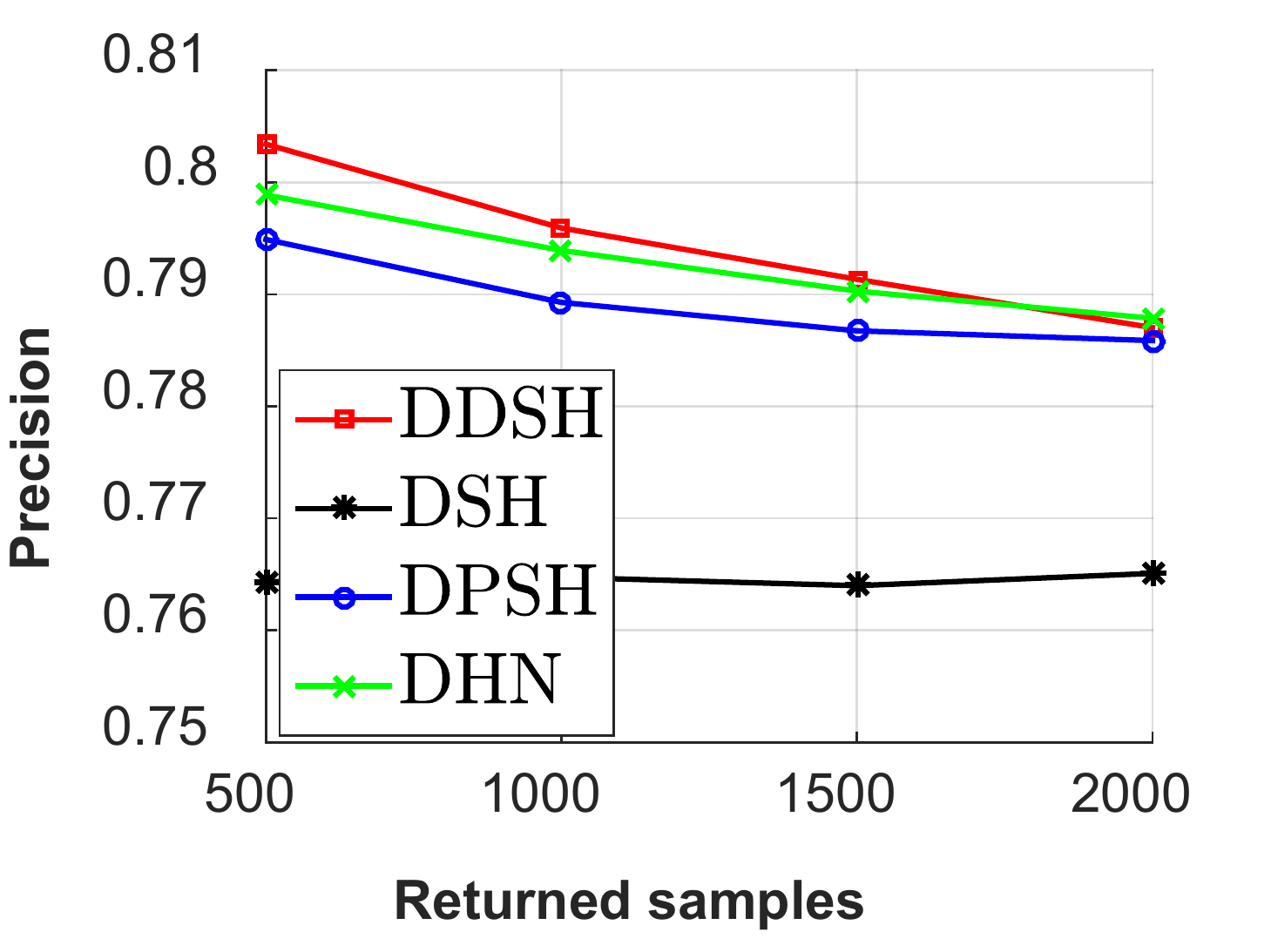}\\
    (i) 12 bits @NUS-WIDE
\end{minipage} &
\begin{minipage}{0.24\linewidth}\centering
    \includegraphics[width=1\textwidth]{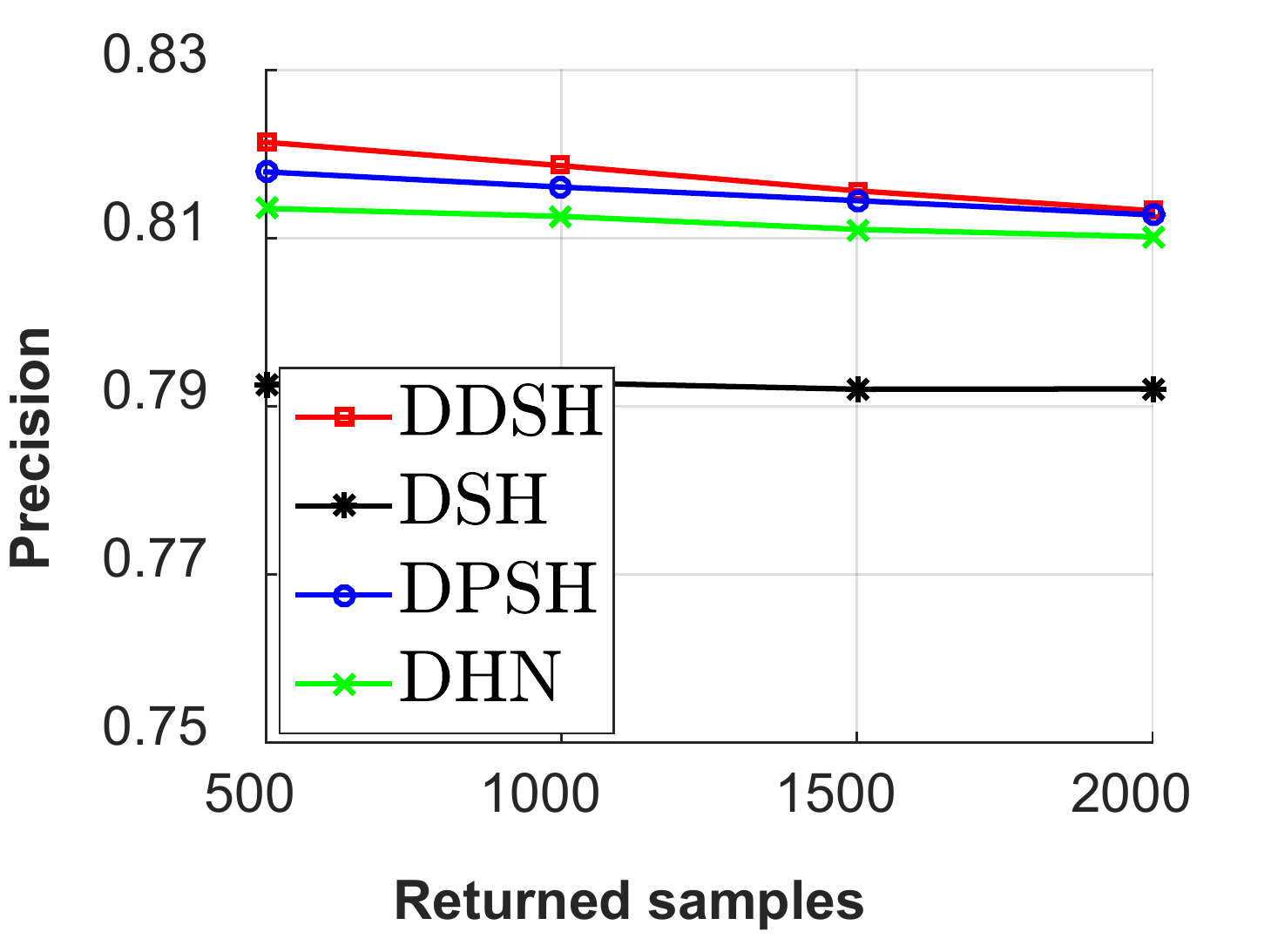}\\
    (j) 24 bits @NUS-WIDE
\end{minipage} &
\begin{minipage}{0.24\linewidth}\centering
    \includegraphics[width=1\textwidth]{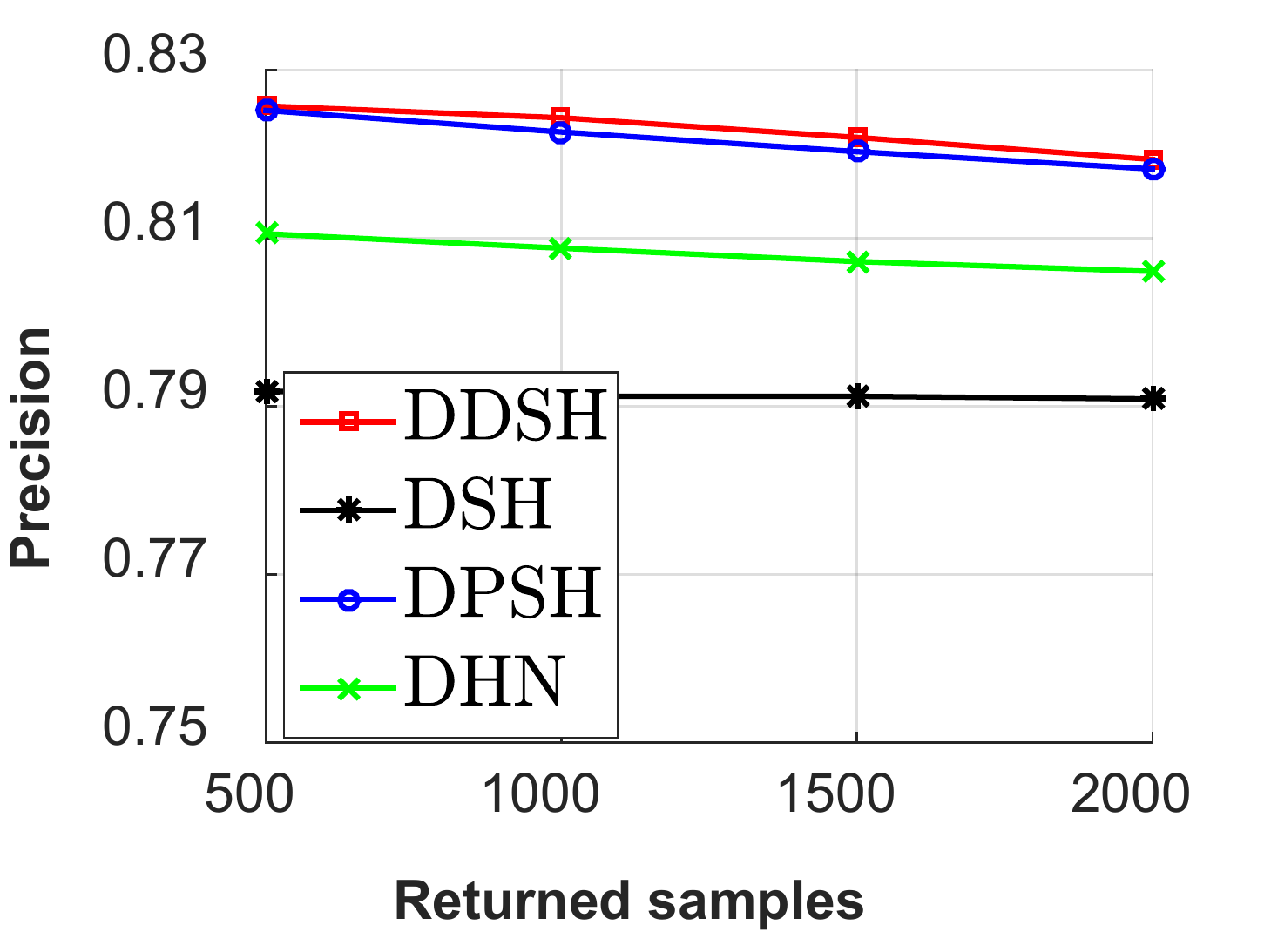}\\
    (k) 32 bits @NUS-WIDE
\end{minipage} &
\begin{minipage}{0.24\linewidth}\centering
    \includegraphics[width=1\textwidth]{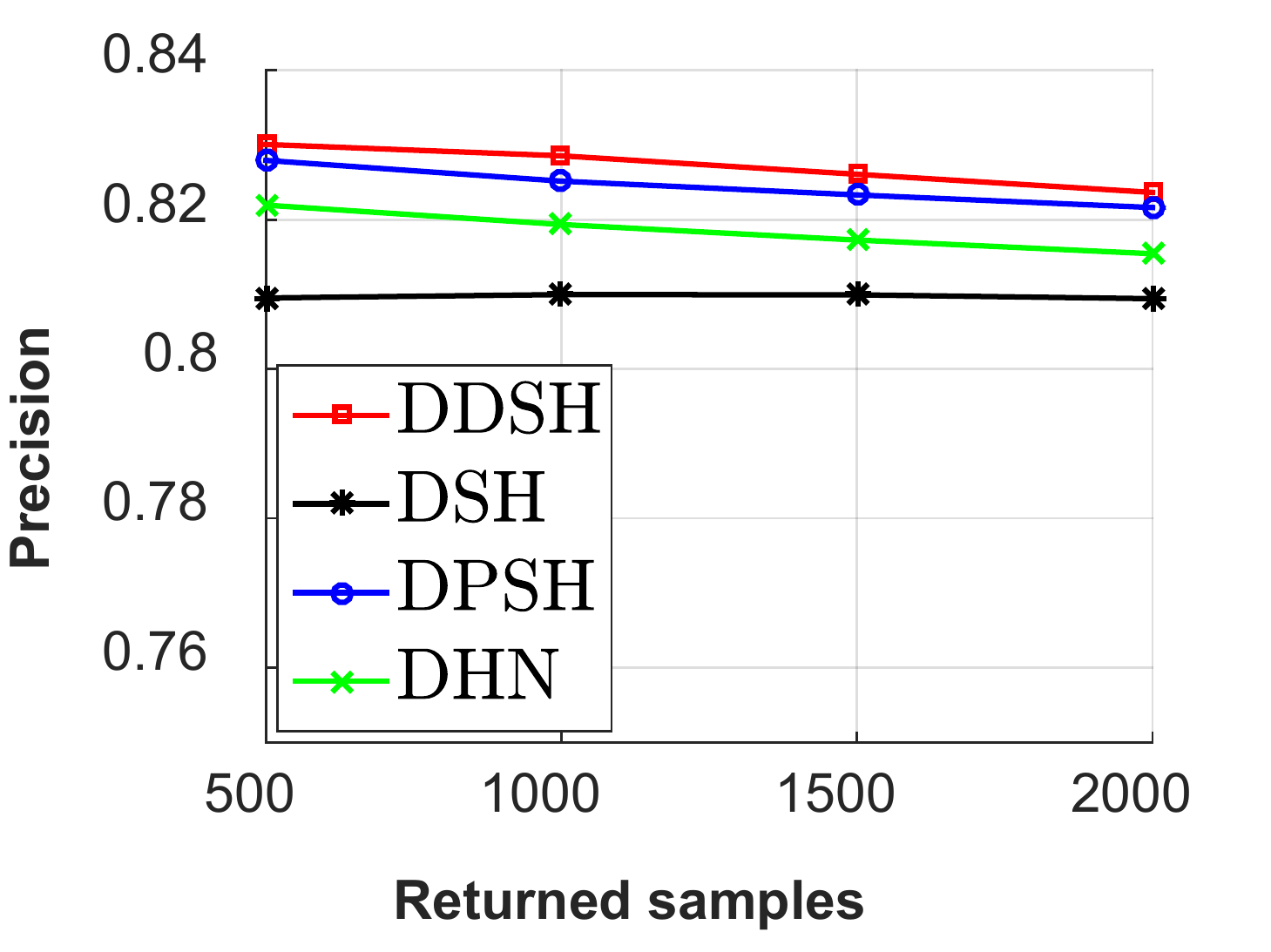}\\
    (l) 48 bits @NUS-WIDE
\end{minipage}
\end{tabular}\vspace*{0pt}
\caption{Performance of top-k precision on three datasets. The four sub-figures in each row are the top-k precision curves for 12 bits, 24 bits, 32 bits and 48 bits respectively.}
\label{fig:top2kpre}
\end{figure*}

In real applications, we might care about top-k retrieval results more than the whole database. Hence we report the \mbox{top-k} precision on three datasets based on the returned \mbox{top-k} samples. In Figure~\ref{fig:top2kpre}, we show the top-k precision for different k on CIFAR-10 dataset, SVHN dataset and NUS-WIDE dataset respectively, where k is the number of returned samples. Again, we can find that DDSH can outperform other deep hashing methods in most cases.

\subsubsection{Hash Lookup Task}
In practice, retrieval with hash lookup can usually achieve constant or  sub-linear search speed in real applications. Recent works like DGH~\cite{DBLP:conf/nips/LiuMKC14} show that discrete hashing can significantly improve hash lookup success rate.

\begin{figure*}[htb]
\centering
\small
\begin{tabular}{c@{ }@{ }c@{ }@{ }c@{ }@{ }c}
\begin{minipage}{0.325\linewidth}\centering
    \includegraphics[width=1\textwidth]{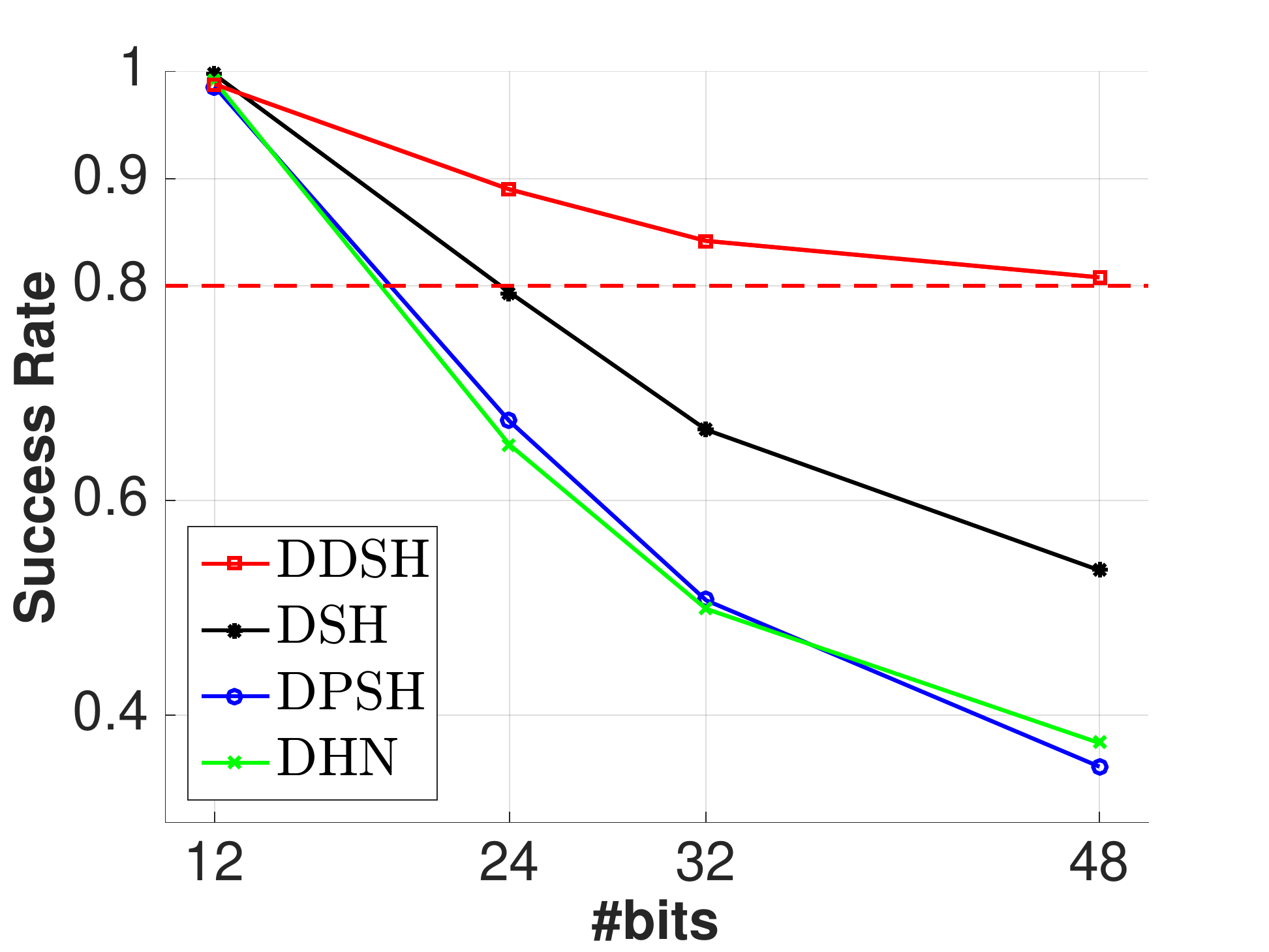}\\
    (a) CIFAR-10 @radius 0
\end{minipage} &
\begin{minipage}{0.325\linewidth}\centering
    \includegraphics[width=1\textwidth]{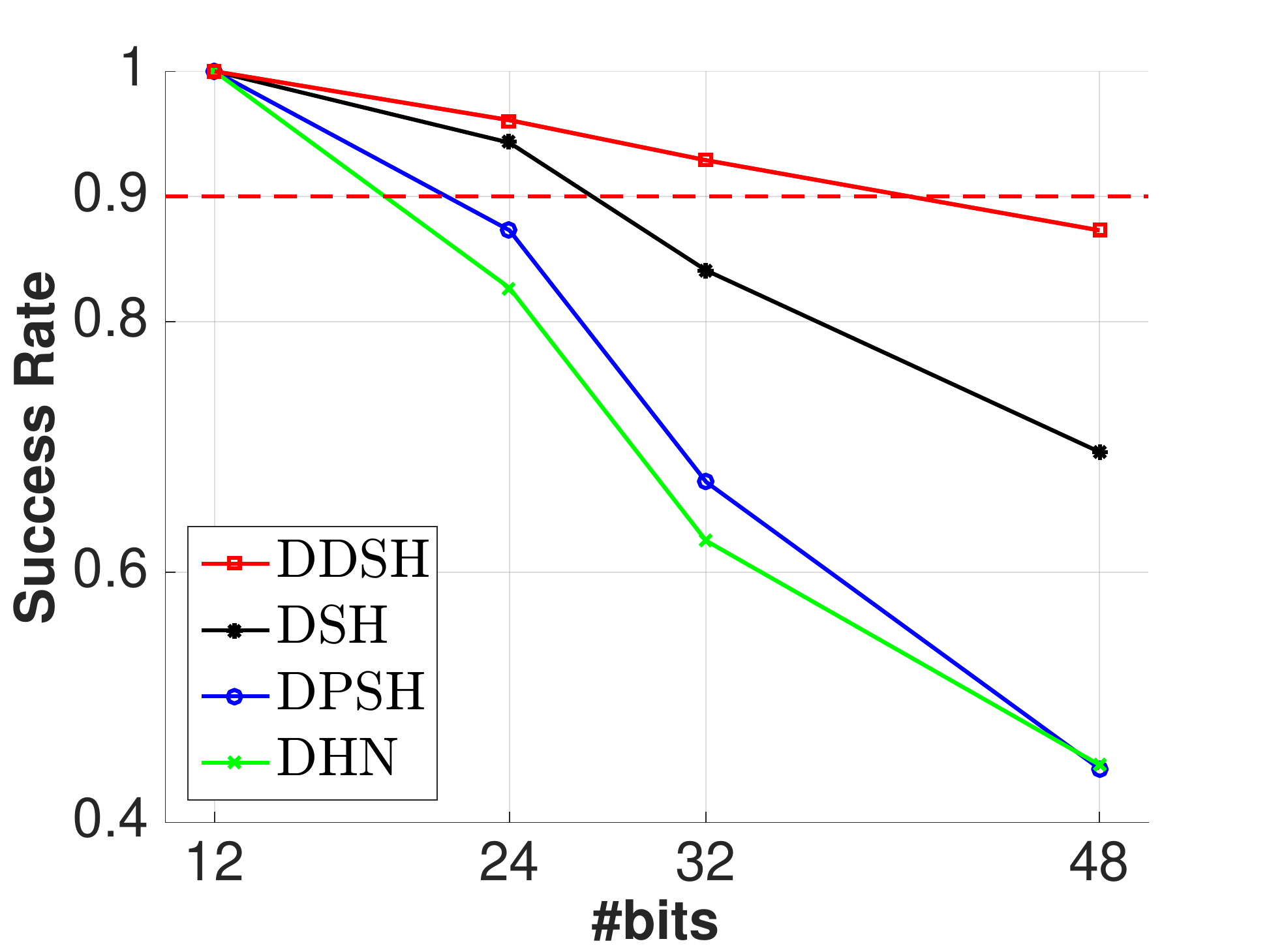}\\
    (b) CIFAR-10 @radius 1
\end{minipage} &
\begin{minipage}{0.325\linewidth}\centering
    \includegraphics[width=1\textwidth]{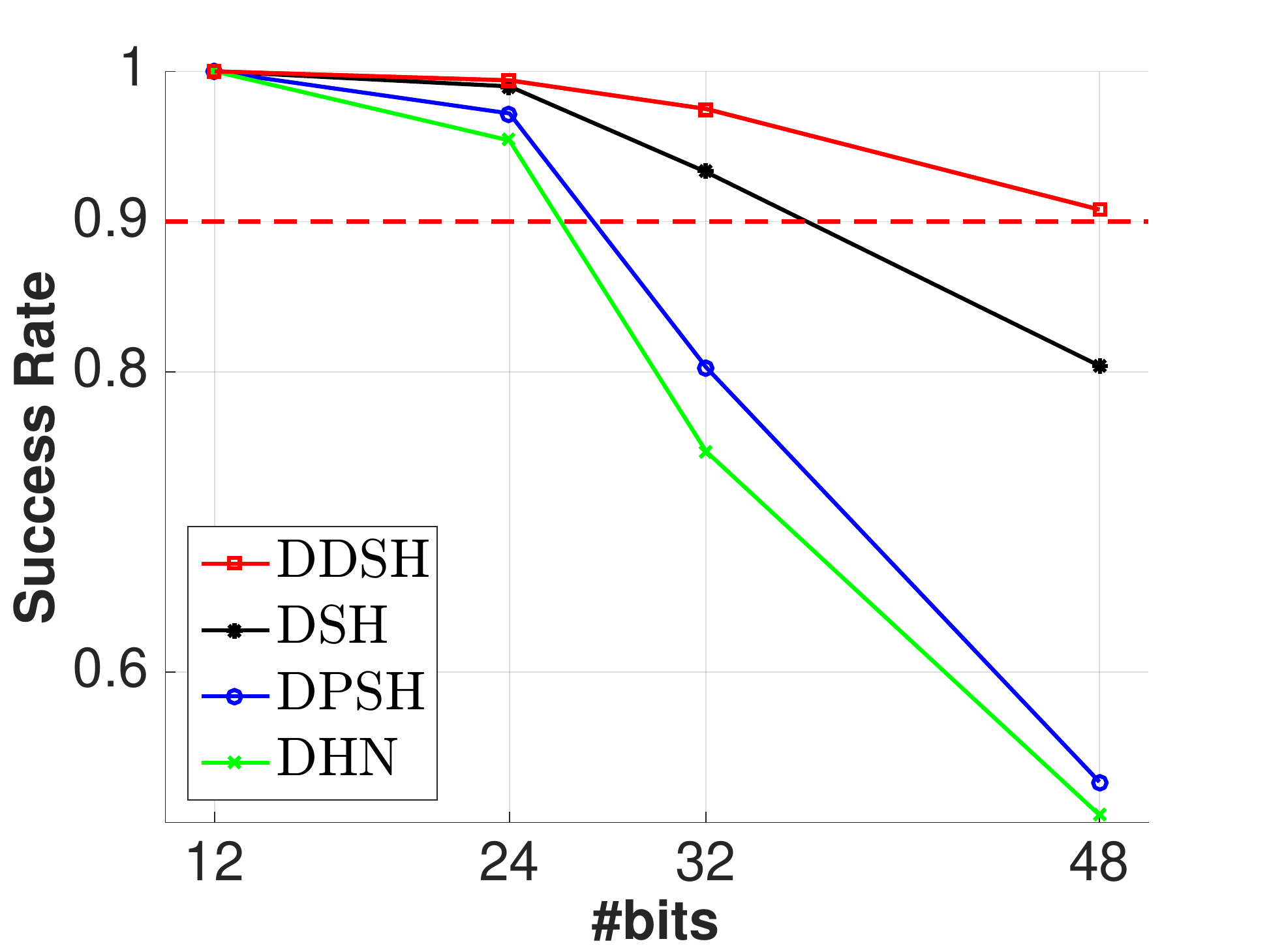}\\
    (c) CIFAR-10 @radius 2
\end{minipage}\vspace{10pt}\\
\begin{minipage}{0.325\linewidth}\centering
    \includegraphics[width=1\textwidth]{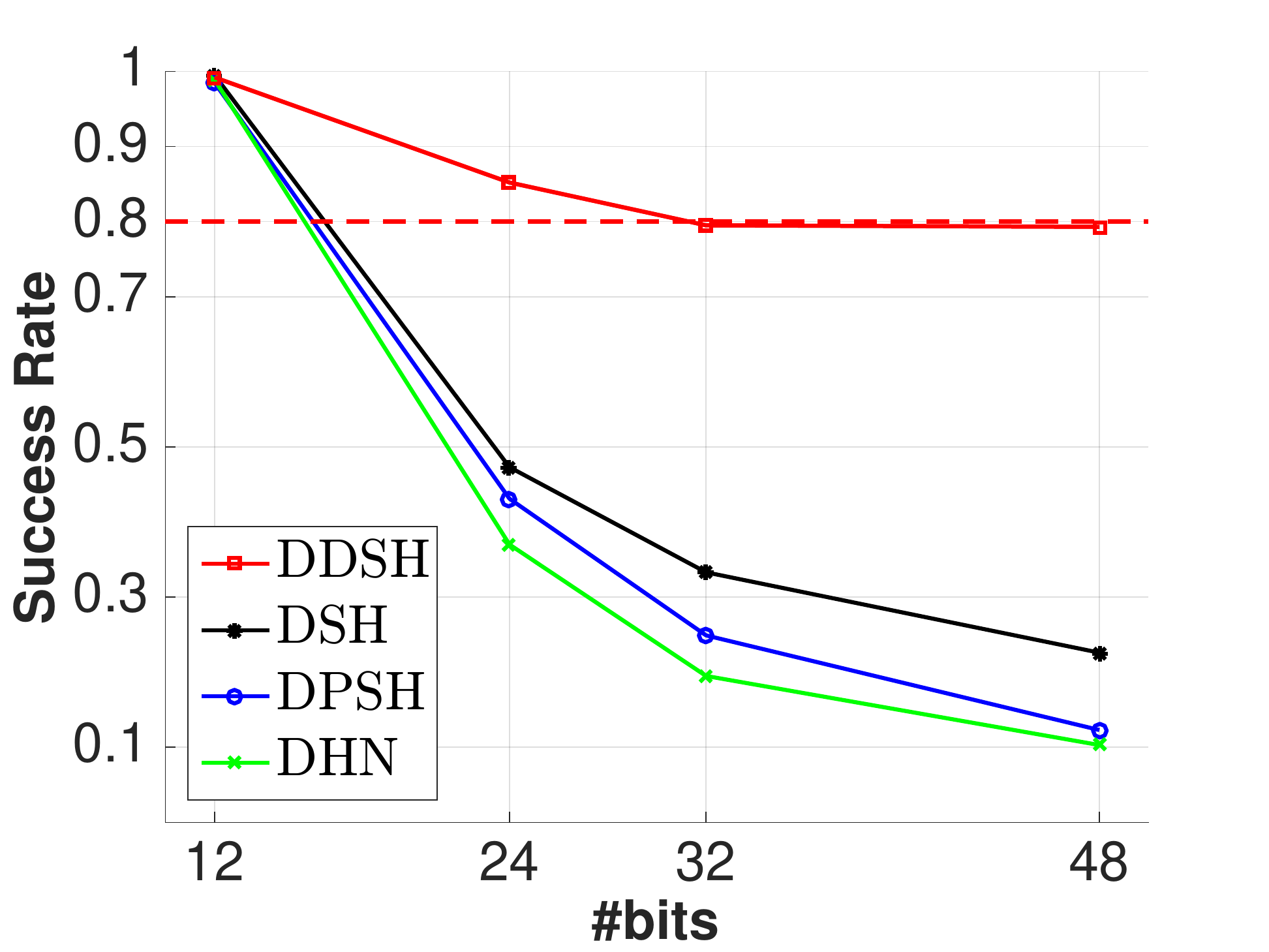}\\
    (d) SVHN @radius 0
\end{minipage} &
\begin{minipage}{0.325\linewidth}\centering
    \includegraphics[width=1\textwidth]{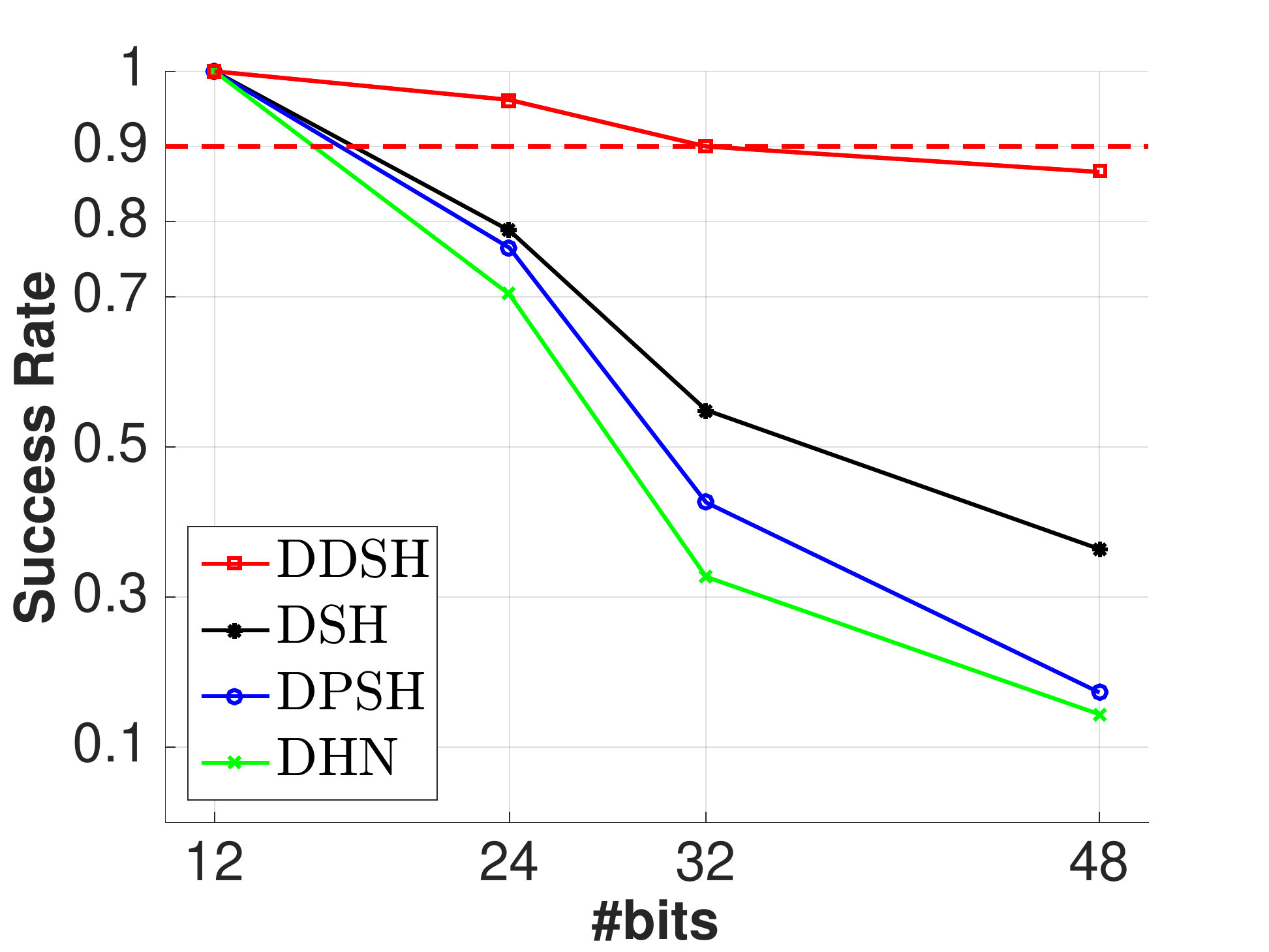}\\
    (e) SVHN @radius 1
\end{minipage} &
\begin{minipage}{0.325\linewidth}\centering
    \includegraphics[width=1\textwidth]{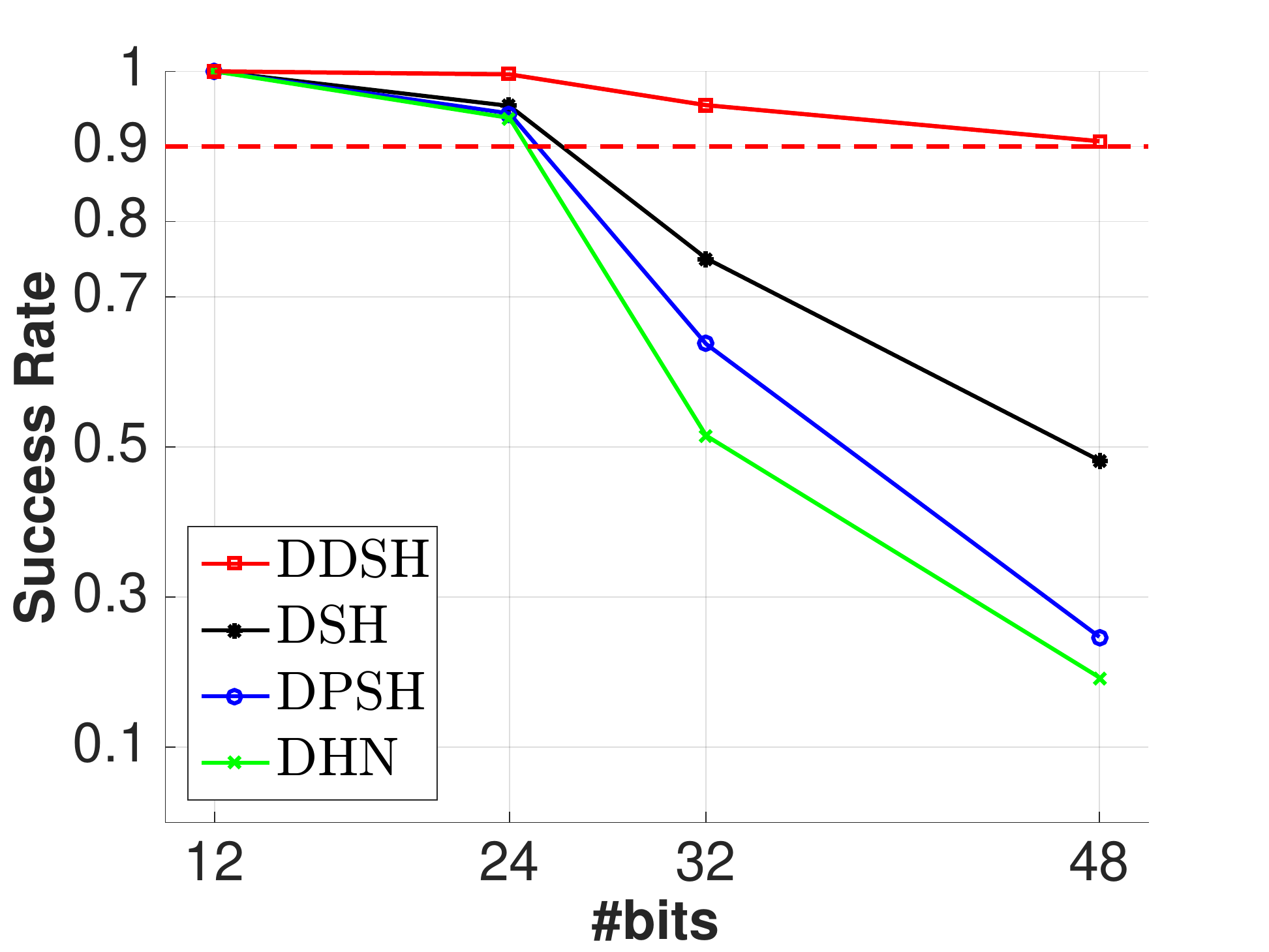}\\
    (f) SVHN @radius 2
\end{minipage}\vspace{10pt}\\
\begin{minipage}{0.325\linewidth}\centering
    \includegraphics[width=1\textwidth]{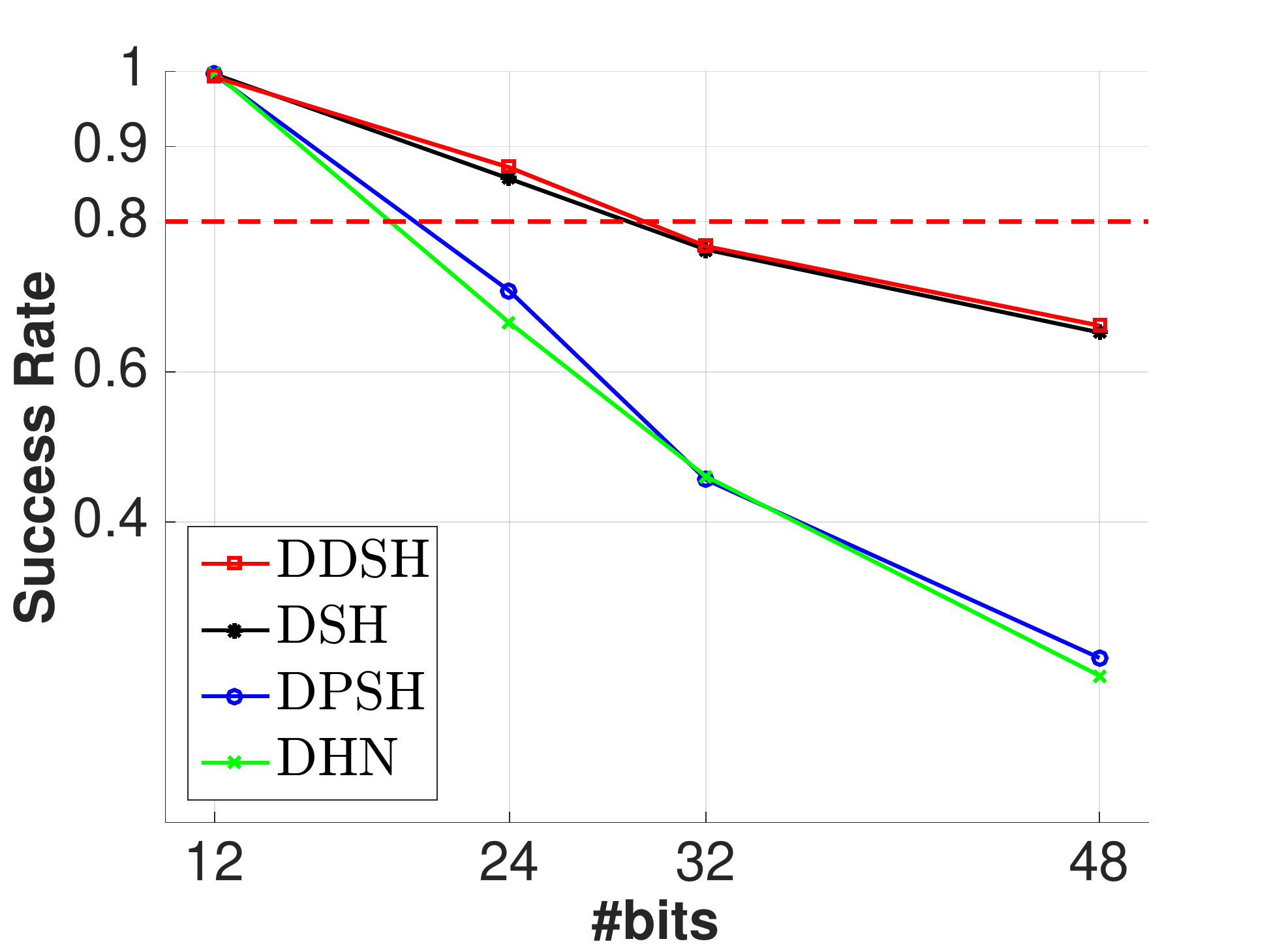}\\
    (g) NUS-WIDE @radius 0
\end{minipage} &
\begin{minipage}{0.325\linewidth}\centering
    \includegraphics[width=1\textwidth]{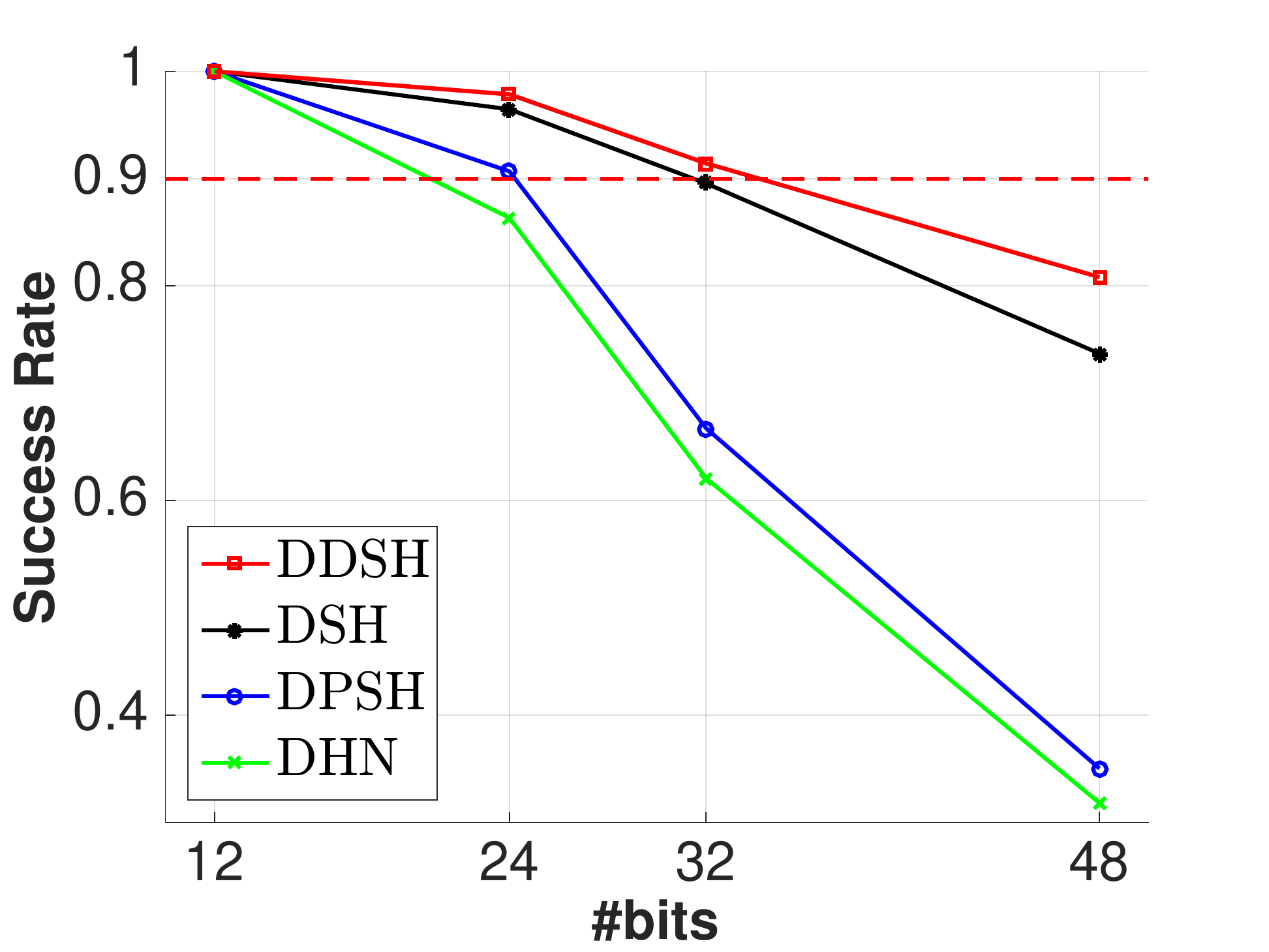}\\
    (h) NUS-WIDE @radius 1
\end{minipage} &
\begin{minipage}{0.325\linewidth}\centering
    \includegraphics[width=1\textwidth]{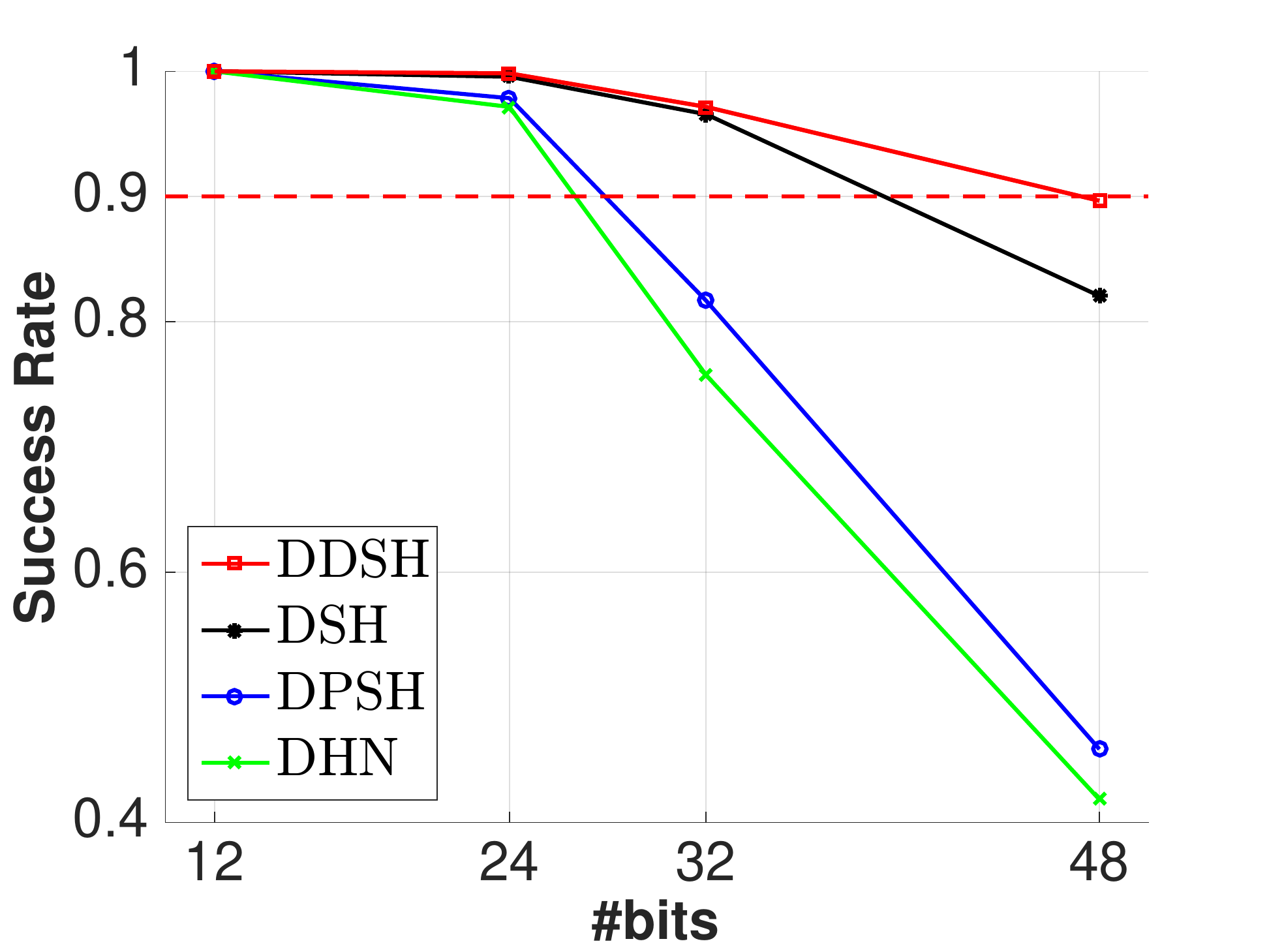}\\
    (i) NUS-WIDE @radius 2
\end{minipage}
\vspace*{0pt}
\end{tabular}
\caption{Hash lookup success rate. Each row includes three sub-figures and presents the hash lookup success rate results on CIFAR-10~((a), (b), (c)), SVHN~((d), (e), (f)) and NUS-WIDE~((g), (h), (i)), respectively.}
\label{fig:hashlookupsucc}
\end{figure*}

In Figure~\ref{fig:hashlookupsucc}, we present the mean hash lookup success rate within Hamming radius 0, 1 and 2 on all three datasets for all deep hashing methods. We can find that DDSH can achieve the best mean hash lookup success rate on three datasets, especially for long codes. Furthermore, the hash lookup success rate of DDSH is nearly above $0.9$ in all cases for Hamming radius 2.



\subsubsection{Further Analysis}
To further demonstrate the effectiveness of utilizing supervised information to directly guide both discrete coding procedure and deep feature learning procedure in the same end-to-end framework, we evaluate several variants of DDSH. These variants include ``DDSH0'', ``COSDISH-Linear'', ``COSDISH-CNN'' and ``DDSH-MAC''.

DDSH0 denotes the variant in which we fix the parameters of the first seven layers of CNN-F in DDSH during training procedure. In other words, DDSH0 can not perform deep feature learning procedure, and all the other parts are exactly the same as those in DDSH. Comparison between DDSH0 and DDSH is to show the importance of deep feature learning.

COSDISH-Linear denotes a variant of COSDISH in which we use linear function rather than boosted decision tree for out-of-sample extension. COSDISH-CNN denotes a variant of COSDISH in which we learn optimal binary codes using COSDISH first, and then we use the CNN-F to approximate the binary codes for out-of-sample extension. Because the discrete coding procedure in DDSH is similar to that in COSDISH, COSDISH-CNN can be considered as a two-stage variant of DDSH where the discrete coding stage is independent of the feature learning stage. The comparison between COSDISH-CNN and DDSH is to show that integrating the discrete coding procedure and deep feature learning procedure into the same framework is important.

DDSH-MAC is a variant of DDSH by using the method of auxiliary coordinates~(MAC) technique in AFFHash~\cite{raziperchikolaei2015learning}. That is to say, we use loss function $\LM_{\text{COSDISH}}(\B^\Gamma,\B^\Omega)+\lambda\Vert\B-\text{tanh}(F(\X;\Theta))\Vert^2_F$ to enhance the feedback between deep feature learning and discrete code learning. Here, $\LM_{\text{COSDISH}}(\cdot)$ is the loss used in COSDISH. DDSH-MAC can integrate the discrete coding procedure and deep feature learning procedure into the same framework. However, the supervised information $S_{ij}$ isn't directly included in the deep feature learning term $\Vert\B-\text{tanh}(F(\X;\Theta))\Vert^2_F$ in DDSH-MAC. That is to say, the supervised information is not directly used to guide the deep feature learning procedure.

The experimental results are shown in Table~\ref{tab:cifar_10_ea_map}. By comparing DDSH to its variants including DDSH0, COSDISH-Linear, COSDISH-CNN and DDSH-MAC, we can find that DDSH can significantly outperform all the other variants. It means that utilizing supervised information to directly guide both discrete coding procedure and deep feature learning procedure in the same end-to-end framework is the key to make DDSH achieve state-of-the-art retrieval performance.

\begin{table}[htb]
\centering
\caption{MAP comparison among variants of DDSH on CIFAR-10. The best accuracy is shown in boldface.}
\label{tab:cifar_10_ea_map}
\begin{tabular}{|c||c|c|c|c|}
 \hline
 \multirow{2}{*}{Method} &
 \multicolumn{4}{c|}{{CIFAR-10}}\\
 \cline{2-5} & 12 bits & 24 bits & 32 bits & 48 bits\\
 \hline \hline
DDSH & {\bf 0.769} & {\bf 0.829} & {\bf 0.835} & {\bf 0.819}\\
\whline
DDSH0 & 0.579 & 0.639 & 0.654 & 0.680\\
\hline
COSDISH-Linear & 0.212 & 0.235 & 0.258 & 0.272\\
\hline
COSDISH-CNN & 0.374 & 0.477 & 0.468 & 0.515\\
\hline
DDSH-MAC & 0.412 & 0.506 & 0.528 & 0.534\\
\hline
 \end{tabular}
\end{table}

\begin{figure}[htb]
\centering
\includegraphics[scale=0.48]{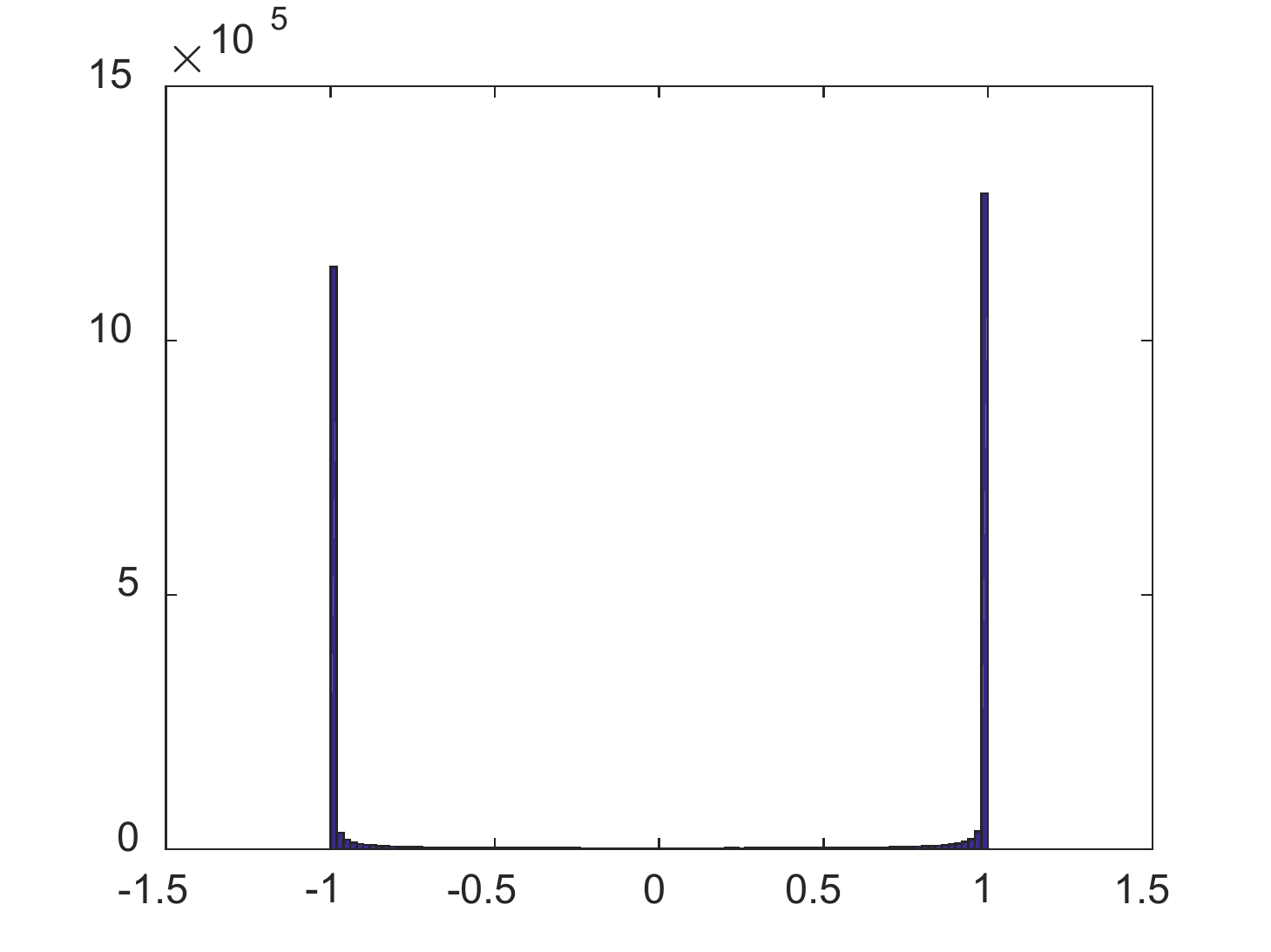}\vspace*{0pt}
\caption{The effect of $\text{tanh}(\cdot)$ approximation on CIFAR-10}
\label{fig:tanh}
\end{figure}

Furthermore, to evaluate the approximation we used when we update the parameter of deep neural network, we report the distribution of the output for the deep neural network. Figure~\ref{fig:tanh} shows the distribution of the output of $\text{tanh}(F(\X;\Theta))$ when we finish the training procedure of DDSH on CIFAR-10 dataset. The x-axis is the $\text{tanh}(F(\X;\Theta))$, and the y-axis is the number of points having the corresponding $\text{tanh}(F(\cdot))$ value. It's easy to see that the $\text{tanh}(\cdot)$ can successfully approximate the $\text{sign}(\cdot)$ function in real applications.

\subsubsection{Case Study}
\begin{figure*}[t]
\centering
\small
\begin{tabular}{c@{ }@{ }c@{ }@{ }c@{ }@{ }c}
\begin{minipage}{0.95\linewidth}\centering
    \includegraphics[width=1\textwidth]{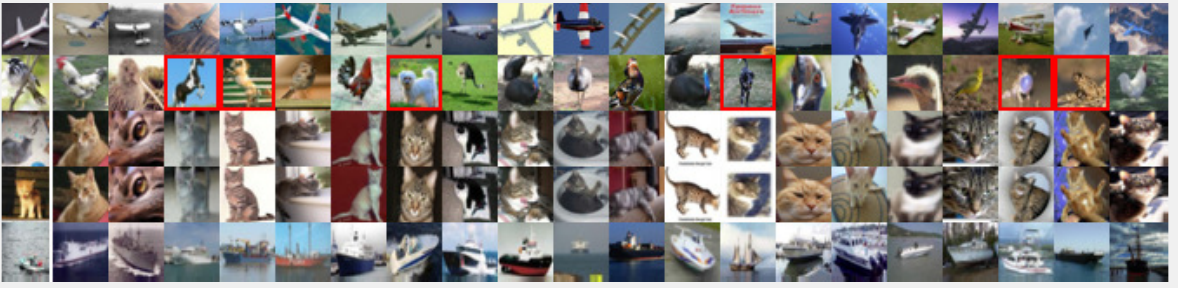}\\
    (a) DDSH @32 bits
\end{minipage}\\
\begin{minipage}{0.95\linewidth}\centering
    \includegraphics[width=1\textwidth]{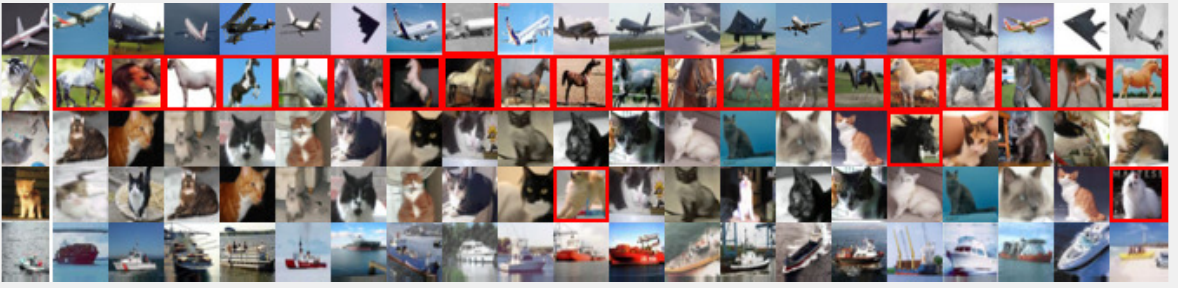}\\
    (b) DSH @32 bits
\end{minipage}\\
\begin{minipage}{0.95\linewidth}\centering
    \includegraphics[width=1\textwidth]{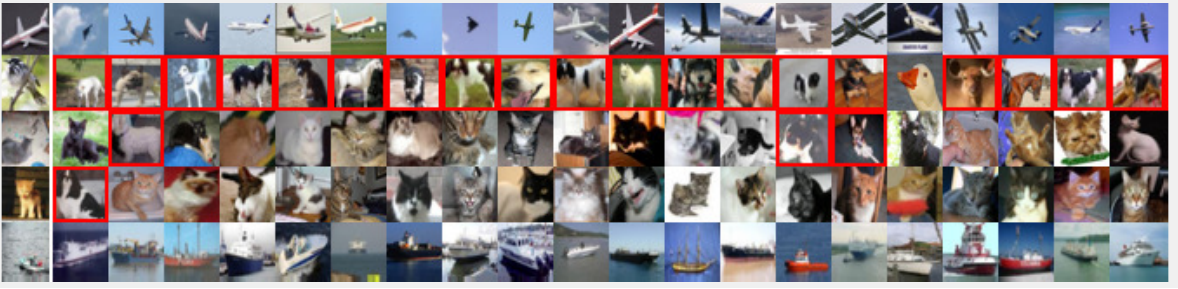}\\
    (c) DPSH @32 bits
\end{minipage}\\
\begin{minipage}{0.95\linewidth}\centering
    \includegraphics[width=1\textwidth]{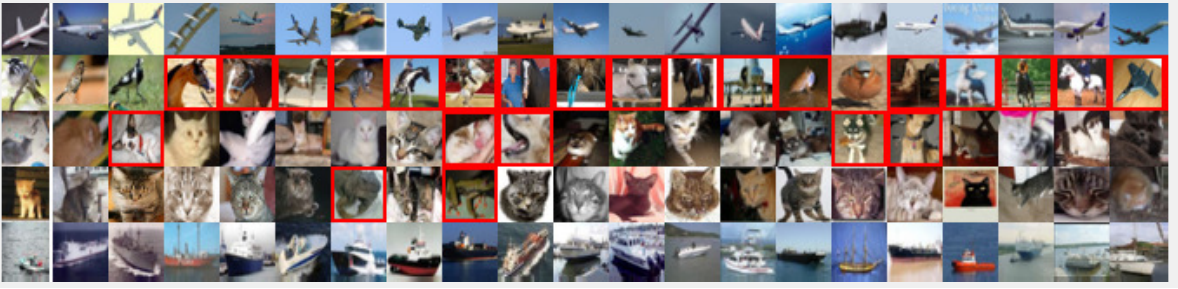}\\
    (d) DHN @32 bits
\end{minipage}\vspace*{2pt}
\end{tabular}
\caption{Case study on {CIFAR-10} with 32 bits. The first column for each sub-figure is queries and the following twenty columns denote the top-20 returned results.
We use red box to denote the wrongly returned results.}
\label{fig:case_study}
\end{figure*}

We randomly sample some queries and return top-20 results for each query as a case study on CIFAR-10 to show the retrieval result intuitively. More specifically, for each given query image, we return top-20 nearest neighbors based on its Hamming distance away from query. Then we use red box to indicate the returned results that are not a ground-truth neighbor for the corresponding query image.

The result is shown in Figure~\ref{fig:case_study}. In each sub-figure, the first column is queries, including an airplane, a bird, two cats and a ship, and the following twenty columns denote the top-20 returned results. We utilize red box to denote the wrongly returned results. It's easy to find that DDSH can achieve better retrieval performance than other deep hashing baselines.

\section{Conclusion}
\label{sec:conc}
In this paper, we propose a novel deep hashing method called deep discrete supervised hashing~(DDSH) with application for image retrieval. On one hand, DDSH adopts a deep neural network to perform deep feature learning from pixels. On the other hand, DDSH also adopts a discrete coding procedure to perform discrete hash code learning. Moreover, DDSH integrates deep feature learning procedure and discrete coding procedure into the same architecture. To the best our knowledge, DDSH is the first deep supervised hashing method which can utilize supervised information to directly guide both discrete coding procedure and deep feature learning procedure in the same end-to-end framework. Experiments on image retrieval applications show that DDSH can significantly outperform other baselines to achieve the state-of-the-art performance.


%

%

\ifCLASSOPTIONcompsoc
\else
\fi

\ifCLASSOPTIONcaptionsoff
  \newpage
\fi



\bibliographystyle{IEEEtran}
\bibliography{IEEEabrv}
%
%
%

%

%
%
%
%
%

\ifCLASSOPTIONcaptionsoff
  \newpage
\fi

\begin{IEEEbiography}[{\includegraphics[width=1in,height=1.25in,clip,keepaspectratio]{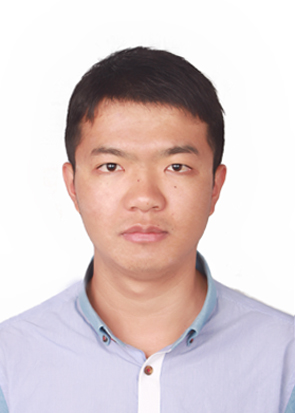}}]{Qing-Yuan Jiang} received the BSc degree in computer science from Nanjing University, China, in 2014. He is currently working toward the PhD degree in the Department of Computer Science and Technology, Nanjing University. His research interests are in machine learning and learning to hash.
\end{IEEEbiography}

\begin{IEEEbiography}[{\includegraphics[width=1in,height=1.25in,clip,keepaspectratio]{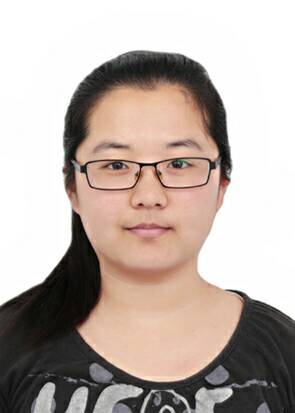}}]{Xue Cui} received the BSc degree in computer science and technology from Chongqing University, China, in 2015. She is currently a Master student in the Department of Computer Science and Technology, Nanjing University. Her research interests mainly include machine learning and data mining.
\end{IEEEbiography}
\begin{IEEEbiography}[{\includegraphics[width=1in,height=1.25in,clip,keepaspectratio]{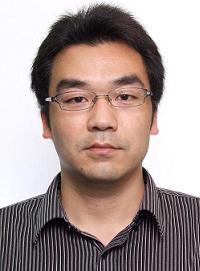}}]{Wu-Jun Li} received the BSc and MEng degrees in computer science from the Nanjing University of China, and the PhD degree in computer science from the Hong Kong University of Science and Technology. He started his academic career as an assistant professor in the Department of Computer Science and Engineering, Shanghai Jiao Tong University. He then joined Nanjing University where he is currently an associate professor in the Department of Computer Science and Technology. His research interests are in machine learning, big data, and artificial intelligence. He is a member of the IEEE.
\end{IEEEbiography}

\end{document}